# Global Photometric Properties of Asteroid (4) Vesta Observed with Dawn Framing Camera

Jian-Yang Li[a,*], Lucille Le Corre[a], Stefan E. Schröder[b], Vishnu Reddy[c, d], Brett W. Denevi[e], Bonnie J. Buratti[f], Stefano Mottola[b], Martin Hoffmann[c], Pablo Gutierrez-Marques[c], Andreas Nathues[c], Christopher T. Russell[g], Carol A. Raymond[f]

[a] Planetary Science Institute, Tucson, AZ 85719, USA

[b] Deutsches Zentrum für Luft- und Raumfahrt (DLR), 12489 Berlin, Germany

[c] Max Planck Institute for Solar System Research, Katlenburg-Lindau, Germany

[d] Department of Space Studies, University of North Dakota, Grand Forks, USA

[e] Johns Hopkins University, Applied Physics Laboratory, Laurel, MD, USA

[f] Jet Propulsion Laboratory, California Institute of Technology, Pasadena, CA, USA

[g] Institute of Geophysics and Planetary Physics, University of California Los Angeles, Los Angeles, CA USA

# Proposed Running Head:

**Photometric Properties of Vesta from Dawn FC**

**Editorial correspondence to:**

Jian-Yang Li

Planetary Science Institute

1700 E. Ft. Lowell Rd., Suite 106

Tucson, AZ 85719

USA

Tel: (571) 488-9999

Email: jyli@psi.edu

**Abstract**

Dawn spacecraft orbited Vesta for more than one year and collected a huge volume of multispectral, high-resolution data in the visible wavelengths with the Framing Camera. We present a detailed disk-integrated and disk-resolved photometric analysis using the Framing Camera images with the Minnaert model and the Hapke model, and report our results about the global photometric properties of Vesta. The photometric properties of Vesta show weak or no dependence on wavelengths, except for the albedo. At 554 nm, the global average geometric albedo of Vesta is 0.38±0.04, and the Bond albedo range is 0.20±0.02. The bolometric Bond albedo is 0.18±0.01. The phase function of Vesta is similar to those of S-type asteroids. Vesta's surface shows a single-peaked albedo distribution with a full-width-half-max ~17% relative to the global average. This width is much smaller than the full range of albedos (from ~0.55× to >2× global average) in localized bright and dark areas of a few tens of km in sizes, and is probably a consequence of significant regolith mixing on the global scale. Rheasilvia basin is ~10% brighter than the global average. The phase reddening of Vesta measured from Dawn Framing Camera images is comparable or slightly stronger than that of Eros as measured by the Near Earth Asteroid Rendezvous mission, but weaker than previous measurements based on ground-based observations of Vesta and laboratory measurements of HED meteorites. The photometric behaviors of Vesta are best described by the Hapke model and the Akimov disk-function, when compared with the Minnaert model, Lommel-Seeliger model, and Lommel-Seeliger-Lambertian model. The traditional approach for photometric correction is validated for Vesta for >99% of its surface where reflectance is within ±30% of global average.

## 1. Introduction

High-resolution images of asteroids acquired from flyby and rendezvous missions have significantly advanced our understanding of the photometric properties of asteroidal surfaces through disk-resolved photometric analyses (e.g., Helfenstein et al., 1994; 1996; Clark et al., 1999; 2002; Kitazato et al., 2008; Spjuth et al., 2012; Magrin et al., 2012). The large range of illumination, viewing, and phase angles available with close-by data that are inaccessible from the ground-based, unresolved or poorly resolved images has significantly improved the modeling of photometric properties. Photometric modeling of the targets represents an essential aspect of the data analysis of all Solar System exploration missions for three reasons. First, photometric properties contain information about the surface physical and mineralogical properties of asteroids. Second, spectroscopic data analyses and interpretations require spectral reflectance data to be compared at a uniform scattering geometry that is the same as laboratory measurements of relevant planetary analogs, and photometric modeling provides such photometric corrections. And third, the planning and design of the acquisition of imaging and spectroscopic data for a spacecraft mission need photometric models to predict the brightness of target objects, both in a disk-integrated sense and a disk-resolved sense, for the purposes of estimating, e.g., exposure times and total durations needed to acquire images, etc.

The Dawn spacecraft approached its first target, Asteroid (4) Vesta, in May 2011 and stayed in Vesta's orbit until its departure in August 2012. The Framing Camera (FC) (Sierks et al., 2011) onboard the spacecraft collected images of Vesta through seven narrow-band filters (~50 nm bandwidth) from 438 nm to 961 nm and one broad-band filter centered at ~700 nm. These images spanned phase angles from a few degrees to ~108º during approach, and pixel sizes from 114 km/pix for the first set of images down to ~0.02 km/pix for images collected at the lowest orbit with a radius of ~500 km. The Visible and Infrared Spectrometer (VIR) obtained a huge amount of spectral maps of Vesta's surface (De Sanctis et al., 2012). In this paper, we report the detailed photometric analysis of the global photometric properties of Vesta using Dawn FC data. During Dawn's stay in orbit around Vesta, the southern hemisphere of Vesta underwent summer solstice to vernal equinox, caused by the tilt of its rotational pole (RA=309.03º, Dec=42.23º, cf. Russell et al., 2012) with respect to its orbital plane by 27.5º. Therefore the photometric properties of Vesta we derived represent the surface south of ~+30º latitude. As will be demonstrated later, the results should be representative of the global photometric properties of Vesta.

Vesta is one of the highest albedo asteroids in the main asteroid belt, with a visual geometric albedo of ~0.38 (Tedesco et al., 2002; Li et al., 2011). The intensive photometric and spectroscopic observations from the ground and Hubble Space Telescope (HST) showed that Vesta has large albedo, color, and spectroscopic heterogeneities on its surface (e.g., Jaumann et al., 1996; Gaffey, 1997; Binzel et al., 1997; Thomas et al., 1997a; Vernazza et al., 2005; Li et al., 2010; Reddy et al., 2010). The albedo and color heterogeneity of Vesta's surface, as well as the mineralogical diversity, have been confirmed by Dawn (Reddy et al., 2012a; De Sanctis et al., 2012; 2013). The shape of Vesta is close to an oblate spheroid, with a difference of <3% between the long- and intermediate-axes (Thomas et al., 1997b; Jaumann et al., 2012), and its rotational lightcurve is therefore dominated by its albedo variations (e.g., Taylor, 1973; Degewij, 1978; Binzel et al., 1997; Li et al., 2010). Despite the intensive studies of Vesta in the past, its photometric properties have never been studied in great detail due to the limitation of scattering angles accessible from the ground. Lagerkvist and Magnusson (1990) reported $G$=0.33 in $V$- band (550 nm) for the phase function of Vesta fitted by the IAU H-G empirical phase function model (Bowell et al., 1989). Hasegawa et al. (2009) derived a $G$=0.21 in $R$-band. Fornasier et al. (2011) derived a $G$=0.27 in $R$-band. Belskaya and Shevchenko (2000) showed that the phase slope and opposition surge of Vesta are comparable with those of S-type asteroids. Helfenstein and Veverka (1989) derived a Hapke model (cf. Hapke, 2012) to fit the disk-integrated phase function of Vesta and derived a single-scattering albedo (SSA) of 0.40, the asymmetry factor of a single-term Henyey-Greenstein (HG) average particle single-scattering phase function (SSPF) -0.30, an amplitude of 1.03 and width of 0.044 for the opposition surge, by assuming a roughness of 20º. Li et al. (2011) combined various observations and derived the geometric albedos of Vesta from 200 nm to 960 nm.

This paper represents an in-depth study of the global photometric properties of Vesta in the visible wavelengths using the spatially resolved, multispectral Dawn FC data, which provide a large range of scattering angles suitable for this type of analysis. The results can be used as a baseline for future studies of the photometric properties of local areas on Vesta, as well as a starting point to apply photometric corrections to Dawn FC data. A similar analysis was carried out by Schröder et al. (2013b), who investigated the performance of various disk-functions to study the distribution of photometric properties over the surface of Vesta as observed through the FC broadband filter. Adopting an exponential phase function, they identified photometric variations associated with crater walls and ejecta.

## 2. Dawn FC data

We used multispectral imaging data collected by Dawn FC during the Approach and Survey phases through all filters in our studies.

The Dawn FC are a pair of identical 1024x1024 pixel imagers (FC1 and FC2) equipped with seven color filters (0.44-0.98 µm) and one panchromatic filter (Sierks et al. 2011) that image the surface of Vesta with an angular resolution of 93 µrad/pixel. Only the FC2 was used to acquire images during the Vesta phase of Dawn mission. A detailed description of the engineering (electronics and optics), operations and basic calibration of the Dawn FC are given in Sierks et al. (2011).

Basic image processing of FC color and clear data is accomplished in three steps with each step producing higher-level data products that are the inputs for subsequent steps. The description of data products from level 0 to level 1b is from Sierks et al. (2011). In the first step, Telemetry Relational Archiving and Processing is a tool that is used to convert the data stream received on the ground (level 0) to the PDS (Planetary Data System) format image products (level 1a) that are of scientific value. Level 1a data contains unprocessed, uncalibrated digital numbers (DNs) from 0 to 16383. In the second step, Calibration Operational Pipeline removes CCD readout smear from level 1a data and applies bias, dark, flat field correction, and the resulting level 1b data are radiometrically calibrated images (Schröder et al., 2013a). The level 1b data have the same PDS-compliant format as the level 1a data.

In addition to the standard calibrations as outlined above, we performed extra calibrations to the images, including stray light removal and fine tune of the absolute radiometric calibration. It is known since shortly after launch that under extended and/or bright sources the back reflection from CCD front surface interferes with the back side of the filters producing stray light patterns, which amount to 5% of the DNs for a typical full frame, but can be as large as 10% in some of the filters with relatively low signal. These patterns can be reasonably estimated from the level 1b images and subtracted to generate level 1c images, which are used in this work. We then adjusted the absolute radiometric calibration of a few filters so that the disk-integrated spectrum of Vesta from FC matches that measured from VIR. The adjustments are ~-6% for F8 filter (438 nm), ~5% for F2 filter (554 nm), and ~4% for F7 filter (652 nm), and no adjustments to other four filters.

The images we used in our photometric analysis are summarized in Table. 1. During the approach phase to Vesta from May 3 to Aug 6, 2011, FC acquired images of Vesta once every several days with the Optical Navigation (ON) imaging sequences, each lasting from half an hour to more than 5 hrs that cover a full rotation of Vesta, mostly through the clear filter (F1). Three imaging sequences, designated as Rotational Characterizations (RC1, RC2, and RC3), acquired color images of Vesta through all 7 color filters covering a full rotation of Vesta. Color images were also acquired in the imaging sequence ON23, right before the start of the next mission phase, the Survey Orbit. Dawn spacecraft stayed in a polar orbit (orbital plane nearly perpendicular to Vesta's equatorial plane) with a low beta-angle (angle between orbital plane and the Sun-Vesta vector) of ~10º from Aug 11 to Aug 29, 2011 before it started to descend to the next lower orbit, the High-Altitude Mapping Orbit (HAMO). The radius of Dawn's Survey orbit is ~3000 km, yielding a pixel size of ~0.28 km/pix at Vesta. The FC collected images through clear filter during the whole Survey phase, and color images through all color filters during 5 out of a total of 7 orbits around Vesta.

[Table. 1]

For images collected from low altitude polar orbits, like those returned by the Moon Mineralogy Mapper ($M^3$), the sub-spacecraft latitude, together with the beta-angle, determines the phase angle of data, usually resulting in a strong trend between the phase angle and latitude in the data. If there are any systematic latitudinal variations in the photometric properties on the surface, then such variations could skew the photometric models if not otherwise considered and accounted for. In addition, because the instrument is usually nadir-pointed to the surface, if the FOV is much smaller than the size of the object, then the local emission angles will be concentrated near low angles, and the accuracy of the photometric model could potentially be affected, particularly for the determination of roughness parameter through limb-darkening properties. Since photometric roughness is best determined from intermediate to high phase angles (Helfenstein, 1988), we consider that the range of emission angles should reach terminator and limb for reliable modeling. The complicated topography on Vesta helps expand the range of local illumination and viewing geometry, but the global distribution of illumination and viewing geometry is still dominated by the configuration of the object's rotation and spacecraft orbit. This is the reason that we decided to use only Approach and Survey images, where the FOV is at least half of Vesta's diameter, to minimize any trend between scattering geometry and latitude. Until ON18, Vesta remained smaller than the FC field-of-view (FOV) with pixel sizes >1 km/pix. The phase angle coverage of those images is from 24º to 108º. Those images are suitable for measuring the disk-integrated phase function of Vesta. For images collected from ON19 to Survey, Vesta is larger than the FOV. We focused on those images for the disk-resolved photometric modeling, where the phase angles range from ~7.7º to ~80º.

Based on the measurement of Vesta's rotational pole by Dawn (Russell et al., 2012), the sub-solar latitude of Vesta was between -21.3º and -27.5º during the Approach and Survey phases. The northern polar area of Vesta, with latitudes >~+50º, was not illuminated in any data we used. Therefore our measurements and modeling only apply to the surface of Vesta at latitudes south of ~30º, covering ~75% of Vesta's surface.  Later images collected when the sub-solar latitude moved just across the equator and illuminated the northern polar area of Vesta indicate the surface is similar to the low- and mid-northern-latitude areas in terms of reflectance and geomorphology.  Therefore, we do not expect much difference between southern photometric properties and the global photometric properties.

## 3.  Disk-integrated phase function

The total flux of reflected light from Vesta is measured from the calibrated images collected through the clear filter during the approach phase before the asteroid fills the FOV of the camera (Table. 1).  The stray light is insignificant in these images and is ignored.  The sky background was measured as the resistant mean of each image excluding the aperture where we measured the total flux of Vesta, and was subtracted out.  The total sky background within the aperture is <0.5% of total Vesta flux for 89% of the images we measured, and <2% for all images, much smaller than the radiometric calibration uncertainty of the camera of ~5% (Schröder et al., 2013a). Given that the signal-to-noise ratio of FC images of Vesta is >200, the uncertainty of the flux measurements is dominated by the absolute radiometric calibration rather than photon noise and sky background.

After normalizing the total fluxes of Vesta to 1 AU in both heliocentric distance and observing range, we converted them to disk-average reflectance by dividing out the expected solar flux through FC clear filter (Schröder et al., 2013a).  The equivalent wavelength of clear filter for reflectance measurement is ~700 nm.  Using the ground-based spectrum of Vesta (Bus and Binzel, 2003), we estimated a factor of ~0.93 to scale our reflectance measurement through FC clear filter to 550 nm, corresponding to the wavelength of the standard Johnson $V$-band filter.

The disk-averaged reflectance measurements were then converted to magnitudes using the apparent solar magnitude at *V*-band, -26.75 (Cox, 1999). Finally, we corrected the magnitude for the changing apparent cross-section of Vesta as seen by Dawn spacecraft during the approach to an equivalent radius of 260 km. This radius corresponds to the same cross-sectional area of Vesta as observed at low sub-observer latitude for a triaxial ellipsoid with dimensions 286.3×278.6×223.2 km (Russell et al., 2012).

We supplemented our disk-integrated photometric data of Vesta with ground-based observations as archived at the PDS Small Bodies Node (SBN) (Lagerkvist and Magnusson, 1995) at phase angles from 1.7º to 26.2º at *V*-band. We also measured the total brightness of Vesta in images acquired by the OSIRIS camera on board Rosetta at 52.5º phase angle (see Fornasier et al., 2011, for a description of the data set) at *V*-band equivalent. No correction to the apparent cross-section of Vesta was applied to ground-based or Rosetta data because they were all collected at sub-observer latitudes between ~±30º, which only resulted in <5% change in the cross-sectional area. The uncertainty of our total magnitude of Vesta measured from FC images is ~0.05-0.10 mag.

Fig. 1 shows the disk-integrated phase function of Vesta by combining the Dawn FC data, ground-based data, and Rosetta data. It is consistent with an IAU H-G phase function model with H=3.2 and G=0.28 at phase angles <80º. Hicks et al. (2013) showed that when this IAU model fit was compared to a composite V-type and Vesta integrated solar phase curve constructed between 0º and 83º, it overpredicted the brightness of Vesta for solar phase angle greater than 55º. Fornasier et al. (2011) derived a G parameter of 0.27±0.01 using the same ground-based observations at Bowell (Lagerkvist et al., 1995) and Rosetta data they processed independently, both of which we included in our analysis. The FC photometric measurements are entirely consistent with the phase function they derived. The discrepancy between the phase function of Vesta and the IAU H-G model at phase angles >80º is not a surprise, because the H-G phase function model was originally developed for main belt asteroids, which can be observed from the ground only at phase angles ∼<30º (Bowell et al., 1989). The phase slope of Vesta for the most linear part of its phase function between 15º and 60º phase angles is 0.0262 mag/deg. The deviations of Vesta's phase function from the linear model are at phase angles <15º and >65º. The Hapke model phase curve shown in Fig. 1 will be discussed in Section 6.2. Based on the disk-integrated phase function of Vesta, its $V$-band geometric albedo is 0.36, consistent with multiple previous measurements from the ground and HST (Tedesco et al., 2002; Schevchenko and Tedesco, 2006; Li et al., 2011; Fornasier et al., 2011).

[Fig. 1]

## 4. Disk-resolved data

The data unit for disk-resolved photometric modeling is the dimensionless quantity, radiance factor $I/F(i, e, \alpha)$ (Hapke, 2012, pp264), where $I$ is the measured intensity of Vesta from each pixel, $\pi F$ is the incident solar flux, $i$ is the local incidence angle, $e$ is the local emission angle, and $\alpha$ is the phase angle. The images were calibrated to $I/F$ with the standard calibration pipeline and the effective solar flux spectrum through the corresponding filter (Schröder et al., 2013a), with additional calibration and processing applied as described in Section 2. The images we used for this analysis have pixel footprints <0.6 km (Table. 1). The images were first binned by a factor of 2 for ON19-ON22 images, and by a factor of 4 for ON23 and Survey data, to an effective spatial sampling of ∼1 km/pix before $I/F$ data extraction. The local scattering angles ($i$, *e*, *α*) were calculated for each binned pixel with the global digital shape model of Vesta (Jaumann et al., 2012; Preusker et al., 2012) using the ISIS software provided by the United States Geological Survey (Anderson et al., 2004; Becker et al., 2012).

Fig. 2 shows the density plot of the *I/F* data we extracted from images taken through the F2 filter.  In phase angle space, the densest concentration of data points is within phase angles 5º-15º and 30º-40º, corresponding to RC3b and RC3 images, respectively.  The phase function model will be over-weighted by the data points at these phase angles.  With the large span in phase angles of these two densely sampled data-swamps, and the large range of available phase angles of all *I/F* data from ~3º to nearly 90º, this overweighting should not pose any bias in the modeling process.  As we will discuss next, the binning process in scattering geometry space before the modeling also helps mitigate the potential problem of uneven weight.  The coverage in (*i*, *e*) space is large and appears to cover the whole range of illumination and viewing geometries. Although the data points are more concentrated at moderate to high incidence angles around 60º and moderate emission angles at 20º, the model of surface brightness distribution will still be robust.  There is a large number of data points with *I/F*<0.02, concentrated at high incidence angles (85º<*i*<105º) and moderate emission angles (30º<*e*<80º).  The incidence angles of these data points suggest that they are from high northern latitude area and/or shadowed areas inside large craters, and likely have non-zero values due both to multiply scattered light from crater walls and the internal scattered light of the camera.  We discarded these low *I/F* data in our modeling because they are not likely valid data points.

[Fig. 2]

In order to examine any trend between scattering geometry and latitude, we generated density plots of the latitudinal distribution of the $I/F$ data (Fig. 3). The trend between latitude and phase angle is weak (Fig. 3a). Low phase angle data are concentrated within low to mid latitude areas between -45º and +25º, and high phase data are near mid- to high latitude in the southern hemisphere. The strong trend between incidence angle and latitude (Fig. 3b) is inevitable unless the object has a high obliquity and the data collection period spans at least half of a revolution of the object around the Sun. We considered that the trend between incidence angle and latitude would have little effects on photometric modeling for two reasons. First, the topography on Vesta broadens the range of incidence angles. And second, the rotation of Vesta will provide a large range of incidence angles for any particular area on Vesta, regardless of the latitude. The range of emission angle covers the full range from 0º to nearly 90º, with a concentration near 20º, dominated by the approach data, in particular RC3 and RC3b for color images (Fig. 3c). With the characteristics of the $I/F$ data we selected, the derived photometric model should represent the averaged photometric properties of the illuminated area on Vesta without been significantly biased by any systematic equator-to-pole photometric variations.

[Fig. 3]

As discussed later, the Rheasilvia basin in southern mid-to-high latitude has an average albedo ~10% higher than the global average (Section 8). Here we estimate the effect of a bright Rheasilvia basin and the absence of data for latitudes >~30º on global photometric modeling. Rheasilvia basin has a radius of ~200 km (Jaumann et al., 2012), accounting for ~15% of the surface. If we assume that the albedo of areas at latitudes >+30º is similar to the rest of the surface except for Rheasilvia basin, then the average albedo we derived from areas south of +30º latitude would be only up to 1% higher than the true global average of Vesta, much lower than the radiometric calibration uncertainty of our data.

Before we model the $I/F$ data, we binned the data points in the scattering geometry space ($i$, $e$, $\alpha$) with 5º bins for all three parameters. The purpose of this step is to decrease the number of total data points to facilitate the numerical optimization. The binning itself is equivalent to an equal-weighted average and should not change the best-fit models from those derived from the unbinned $I/F$ data. The binning will also partly remove the uneven distribution of the I/F data in scattering geometry space (Figs. 2 and 3). Finally we discarded all binned data points with $i$>80º or $e$>80º in our modeling processes to avoid any possible large uncertainty in the calculated scattering geometry and the alignment with images, and to avoid the limitations of photometric models at extreme scattering geometries.

## 5. Minnaert model

The Minnaert model (Minnaert, 1941) is an empirical model that describes the dependence of reflectance with respect to local scattering geometry, $i$ and $e$. Under the Minnaert model, the radiance factor $I/F$ of a surface is described as

$$I/F(i, e, \alpha) = A(\alpha)\cos^{k(\alpha)}(i)\cos^{k(\alpha)-1}(e) \qquad (1)$$

In this formula, $A(\alpha)$ is commonly termed the Minnaert albedo, and is a function of the phase angle. The phase angle dependence of Minnaert albedo is equivalent to a surface phase function. An empirical model can then be applied to describe it. In our modeling, we adopted a linear function in magnitude with a phase slope parameter, $\beta$, in mag/deg, and a Minnaert albedo at opposition, $A_0$, which is equivalent to normal albedo, to describe $A(\alpha)$,

$$A(\alpha) = A_0 10^{0.4\beta\alpha} \qquad\qquad (2)$$

In Eq. (1), $k(\alpha)$ is the Minnaert limb-darkening parameter, also dependent on phase angle. For dark surfaces such as those of the Moon (Helfenstein and Veverka, 1987) and comets (Li et al., 2009; 2013b), it is close to 0.5 near opposition, indicating a flat brightness distribution across the disk at low phase angles (McEwen, 1991). Previous work showed that for cometary surfaces (Li et al., 2009; 201b), the Minnaert $k$ parameter can be described by a linear function with respect to phase angle with a slope, $b$, in deg$^{-1}$ and a parameter, $k_0$, at opposition, as suggested by McEwen (1991),

$$k(\alpha) = k_0 + b\alpha \qquad\qquad (3)$$

In our modeling process, we fit the $I/F$ data for each color filter to derive the Minnaert $A$ and $k$ parameters for each 5º bin in phase angle. Then we fit both parameters with respect to phase angles with the simple log-linear and linear models, respectively, as described above for each color filter.

The Minnaert model parameters for F2 filter data are plotted in Fig. 4 as an example of the model processes of all filters. The Minnaert $A(\alpha)$ generally follows the log-linear phase function model. The statistical uncertainty for each $A$ is tiny (error bars are smaller than the plotted symbols), but the scatter of $A$ around the best-fit phase function model curve suggests that their actual 1-$\sigma$ uncertainties are probably around ±0.01 (Fig. 4). The deviations of measurements from the log-linear model might also be affected by the large albedo variations across the surface of Vesta, given the slight trend between latitude and phase angle (Fig. 3a). At phase angles <10º, an opposition surge is evident, but we did not include it in our model.

The Minnaert $k(\alpha)$ for Vesta grossly follows a linear relationship with phase angle as well (Fig. 4). At phase angles >65°, the trend shows signs of flattening out at 0.75 - 0.80. This flattening trend is evident in all color filter data, but not the clear data. At zero phase angle, the modeled Minnaert $k$ parameter is ~0.53 at all color filter wavelengths, suggesting a relatively flat disk of Vesta near opposition, similar to that of the Moon (Helfenstein and Veverka, 1987). The overall root-mean-square scatter (or RMS, the square root of the mean of total squared difference between model and data) is 6-8% for color data, and 5% for clear data.

[Fig. 4]

With the small model scatter, the Minnaert model appears to be able to describe the photometric behavior of Vesta's surface sufficiently well. Fig. 5 shows the model goodness assessment. The overall scatter is reasonably small (Fig. 5a). There is no systematic bias of the model with respect to scattering geometry ($i$, $e$, or $\alpha$) (Fig. 5b). The model scatter is almost uniform along both $i$ and $e$ except for $e$>75°, but significantly increases at phase angles >45°. Similar behavior has also been noticed for cometary surfaces of Wild 2 and Tempel 1 fitted by Minnaert models, and attributed to the limitation of the Minnaert model (Li et al., 2009; 2012).

[Fig. 5]

The Minnaert model results at all wavelengths, including those measured through the clear filter, are shown in Fig. 6. The $A_0$ at opposition is close to the geometric albedo, given the nearly flat brightness distribution of the surface of Vesta at low phase angle. However, since we did not include the opposition surge, the values of $A_0$ are slightly lower than the geometric albedo. Nevertheless, the shape of the $A_0$ spectrum, excluding the value from clear filter, should mimic the reflectance spectrum of Vesta (Fig. 6a). The value derived from clear filter data is expected to be below the spectrum of Vesta as measured through narrowband color filters, because the broadband filter results in an average reflectance over its whole bandpass weighted by the product of solar spectrum and system transmissivity. Our calculation suggested good quantitative consistency between the measurements from color filters and the clear filter.

[Fig. 6]

The slope of the Minnaert albedo $A(\alpha)$ is a measure of the slope of the phase function (note that this phase function is a surface phase function, different from the SSPF and from the disk-integrated phase function) (Fig. 6b). The slope appears to be slightly higher within the 1-μm mafic band, indicating a slightly steeper phase function, consistent with the observations of deeper absorption bands with higher phase angles (Reddy et al., 2012b; Sanchez et al., 2012). A slight trend of the slope with respect to wavelengths between 554 nm and 750 nm is also present, consistent with general phase reddening observed for Vesta from the ground and for HED meteorites (Reddy et al., 2012b). We will return to the discussions of phase reddening in Section 9.2.

The Minnaert $k$ parameters at zero phase angle, $k_0$, concentrate within a small numerical range between 0.532 and 0.545 (Fig. 6c). The $k_0$ parameter of Vesta is similar to that of the lunar surface (Helfenstein and Veverka, 1987) and the surfaces of Wild 2 (Li et al., 2009) and Tempel 1 (Li et al., 2013b) as well. Buratti (1984) showed that for a range of albedos on the icy Saturnian moons, their surfaces tended to be "lunar-like" until they reached a geometric albedo of about 0.6; our result for Vesta is consistent with this trend. The slopes of $k$ parameters with respect to phase angle are between 0.0035 and 0.0038. There is no obvious trend evident (Fig. 6d). The physical interpretation of the Minnaert $k$ parameter is not clear. But since it is dominated by the limb-to-terminator reflectance variations, this parameter is affected by both the optical properties of the surface, such as transparency and mechanical properties such as roughness. The former could be wavelength-dependent, and the latter should usually be wavelength-independent for a spectrally grey object. The lack of any trend of either the extrapolated Minnaert $k$ parameter at opposition, or the slope with respect to phase angle at all wavelengths studied suggests that any variations of the optical properties of Vesta at visible wavelengths are not significant enough to cause appreciable wavelength-dependency in $k$.

## 6. Hapke model

### 6.1 Hapke model

During the continuous improvement of Hapke model since it was first introduced in the 1980s, a number of variations for this model with different numbers of model parameters and formalism for its components have been developed. In our modeling, we adopted a five-parameter version (cf. Hapke, 2012), including the SSA, an asymmetry factor, $g$, of the average particle SSPF with a single-term Henyey-Greenstein (HG) function, a macroscopic roughness parameter, $\bar{\theta}$, and the amplitude, $B_0$, and width, $h$, of the opposition surge. This choice was determined by the characteristics of the available data, the attempt to keep the minimum number of model parameters included, and our experiments with different versions in the model fitting process. Due to the lack of data at high phase angles, we cannot model any forward scattering lobe, and therefore a double-term HG function for the SSPF is not necessary. The lack of sufficient data at small phase angles within the possible opposition surge does not allow us to constrain the opposition parameters. Therefore, including the coherent-backscattering opposition effect (CBOE) in our model achieves nothing other than adding more complexity to the model.

We only included the shadow-hiding opposition effect (SHOE) in our model, and will discuss the possible effect of CBOE in the interpretations of the results. We experimented with both the isotropic multiple scattering model as originally introduced in Hapke model (cf. Hapke, 2012) and the anisotropic multiple scattering model in a later improvement (Hapke, 2002) in our modeling. Rather than improving the model fit, the use of anisotropic multiple scattering returned worse model residuals than the isotropic multiple scattering model. Thus we decided to choose isotropic multiple scattering model in our analysis. Finally, we attempted to include the newly introduced porosity parameter (Hapke, 2008) in our modeling, but could not retrieve any meaningful results for the porosity parameter. All of the best-fit values of the filling factor approach zero, equivalent to the form of Hapke model without porosity included. This might be due to the fact that the derivation of porosity primarily relies on multiple scattering in Hapke's model, while the multiple scattering on Vesta (15-30%, see Section 9.1) is relatively weak compared to bright, icy objects. We excluded the porosity parameter in our analysis.

The basic procedure for Hapke model fitting is similar to that used by Li et al. (2004; 2006). We used the MPFIT package found in the Markwardt IDL library (Markwardt, 2008) to search for the best-fit parameter set that minimizes the $\chi^2$, defined as

$$\chi^2 = \sum [r(i, e, \alpha) - r_{measured}]^2 \qquad (4)$$

where $r(i, e, \alpha)$ is the model $I/F$ and $r_{measured}$ is the measured $I/F$, and the sum is over all data points. Alternatively, if the measurement uncertainty, $\sigma$, is available for all data points, then $\chi^2$ can be defined as

$$\chi^2 = \sum \left[ \frac{r(i, e, \alpha) - r_{measured}}{\sigma} \right]^2 \qquad (5)$$

MPFIT is an implementation of a rugged minimization algorithm to provide a robust and relatively fast search in parameter space for non-linear models. To avoid landing on a local minimum in $\chi^2$ space due to the highly coupled nature of Hapke model parameters, we performed the least-squares fit with at least 100 randomly generated initial parameter sets for each filter to ensure the model converges to the true global minimum. We also tried an unweighted fit for the binned $I/F$ data, and weighted fit with the square root of $I/F$ data representing the relative significance of photon errors in the modeling. The modeled parameters for the two cases are within 1% from each other for SSA, 3% for $g$ and $\bar{\theta}$, and 20% for $B_0$ and $h$, which are the worst constrained parameters as shall be discussed. The difference of the model parameters from the weighted and unweighted fits is well within the model uncertainties. Therefore we will arbitrarily base our discussions on the unweighted fit results. We will present the modeling process and results for clear filter data in Section 6.2, color filter data in Section 6.3, and discuss the uncertainty of model fitting in Section 6.4.

*6.2 Clear filter*

Approach images were collected only through the clear filter, except for the three RCs, of which RC3 and RC3b had Vesta larger than the FOV. The disk-integrated phase function is therefore available only for the clear filter but not any color filters. For this reason, we first focused on clear filter data with both disk-integrated and disk-resolved modeling. The disk-integrated phase function covers larger range of phase angles up to 108° than the disk-resolved data. However, the roughness parameter and the SSPF have similar effects on a disk-integrated phase function in the Hapke model, in that both a rougher surface and a more backscattering SSPF could cause a steeper disk-integrated phase function. Therefore the disk-resolved, limb-to-terminator reflectance characteristics provide the best information to break the ambiguity and to retrieve the surface roughness (Helfenstein, 1988). For this reason, we decided to start with disk-resolved data, and then use the best-fit roughness parameter to fit the disk-integrated phase function, and compare the results for consistency and validation.

Because of the lack of sufficient data at low phase angles within the possible opposition surge from either disk-integrated phase function and disk-resolved data, the opposition parameters cannot be well constrained. Given the coupling effects of opposition surge and SSPF at small phase angle, the model results of opposition parameter inevitably affect the determination of the $g$-parameter of SSPF. We started with the opposition parameters found in the literature, $B_0$=1.03 and $h$=0.04 (Helfenstein and Veverka, 1989), and explored the modeling of opposition parameters. First, we fixed both $B_0$=1.03 and $h$=0.04 and fitted other three parameters for disk-resolved $I/F$ data through clear filter. Then we fixed $B_0$=1.03 and fitted other four parameters. The best-fit $h$ parameter is 0.076. Finally, we left all parameters free in the fit and derived $B_0$=1.66 and $h$=0.076. The fitted parameters for all cases are listed in Table. 2.

[Table. 2]

It is interesting to note that $h$=0.076 is the best-fit value both for $B_0$=1.03 and for the best-fit $B_0$ of 1.66. For those two cases, only the $g$-parameter is significantly different, and both the SSA and roughness parameter are close to each other. The model RMS is just slightly smaller for the case with all parameters free than other cases. This fact suggests that our data cannot constrain the $B_0$ parameter. But the uncertainties in opposition parameters only strongly affect the fitted SSPF, and both SSA and roughness can still be retrieved reasonably well. The best-fit $B_0$ is greater than unity, contrary to the model assumptions that the amplitude of SHOE should be no larger than unity. This phenomenon is not unique to Vesta. Many Solar System objects, especially dark ones, have their best-fit $B_0$ parameters higher than unity when only the SHOE is included in the Hapke model. A good example is Asteroid (253) Mathilde, which has its opposition surge well observed, and a $B_0$ parameter of 3.18±1.0 (Clark et al., 1999). Other examples include Asteroid (951) Gaspra, (243) Ida (Helfenstein et al., 1994; 1996), and (433) Eros (Clark et al., 2002; Li et al., 2004). The widely accepted explanation is that the opposition surge could also contain CBOE, which causes the total amplitude to exceed unity. We could not pursue the separation of CBOE from SHOE for Vesta because of the insufficient data at low phase angles.

In order to check the consistency between disk-resolved model parameters and the observed disk-integrated phase function, we over-plotted the three disk-resolved models with observed disk-integrated phase function of Vesta in Fig. 7. First, the three different models result in similar disk-integrated phase functions. The largest differences appear within low phase angles <15º where the data are limited. Second, all disk-resolved models provide consistent fit to the observed disk-integrated phase function.

[Fig. 7]

For the disk-integrated phase function, we fixed the roughness parameter as the best-fit value of 18º derived from the disk-resolved modeling with both clear and color filter data (Section 6.3), and experimented with four cases: 1. Fixed both $B_0$=1.03 and $h$=0.04; 2. Fixed $B_0$=1.03; 3. Fixed $h$=0.054, the best-fit value from case 2; and 4. Set all parameters free. The results are listed in Table. 3. The best-fit $h$, when $B_0$ is fixed to 1.0, is 0.054, slightly different from the value fitted from disk-resolved data. The best-fit $B_0$, with $h$ fixed at 0.054, is 1.77, also just slightly different from the value fitted from disk-resolved modeling. However, when all parameters are set free, the best-fit results in $B_0$=2.6, $h$=0.12. We consider these values not as well constrained as those from disk-resolved modeling, because the disk-integrated photometric data at low phase angles are all from previous ground-based observations (Lagerkvist and Magnusson, 1995). Visual inspection shows that some data are systematically offset from others, due possibly to the uncertainty in their radiometric calibrations. The disk-resolved data are all derived from FC images that have been systematically calibrated.

[Table. 3]

Despite the different best-fit values of parameters in all four cases, their associated disk-integrated phase functions all fit the observed phase function of Vesta well, and similar to each other (Fig. 8). The largest differences, similar to the disk-resolved modeling, appear at phase angles <15º, where the opposition parameters dominate the shape of the phase function.

[Fig. 8]

Comparing the $g$-parameter fitted from disk-resolved data (Table. 2) and disk-integrated data (Table. 3), we can see that the agreement is reasonably good for corresponding cases, suggesting that both techniques (disk-integrated vs. resolved) returned reliable results. Note that the albedo values listed in both tables, and the derived geometric albedos and Bond albedos as well, are systematically different for the corresponding cases. This is due to the scaling with a factor of 0.93 that we applied to convert the reflectance measured from FC clear filter at an effective wavelength of 700 nm to $V$-band equivalent in order to combine then with ground-based and Rosetta data for the disk-integrated case. Taking this scaling factor into account, one will find that the consistency for SSAs and geometric albedos is within 5%. The residual difference in geometric albedo is due mainly to the model uncertainty in opposition parameters. The consistency between various models of Bond albedo, which is dominated by data at moderate phase angles where we have good coverage, is within 0.4%. The best-fit absolute magnitude of Vesta in *V*-band is 3.27±0.07, corresponding to a geometric albedo of 0.34±0.02, consistent with the IAU H-G phase function model results (Section 3).

From this study, we also explored the effects of the uncertainties of modeled opposition parameters on the fit of other parameters. As mentioned earlier, the SSPF is most affected by the uncertain opposition surge. For the same set of reflectance data to be modeled, a stronger opposition surge (greater amplitude and width) will account for some slope of the surface phase function, and result in a modeled SSPF that is less backscattering. The SSA is slightly decreased as a result of stronger opposition surge. This is because the SSA is an integrated quantity of a single particle over all $4\pi$ solid angle. Although the SSPF itself is normalized to unity for the integration over the full range of phase angles, a stronger opposition surge results in higher scattered light. Thus a smaller SSA is required to yield the same bidirectional reflectance. Quantitatively, based on all the cases we explored for disk-resolved modeling, changing the $B_0$ from 1.03 to 1.66 and $h$ from 0.04 to 0.076 results in a change of SSA by 0.02 and $g$ by 0.06. From Table. 2 and Table. 3 (except for Case 4 for disk-integrated phase function model), we consider that the uncertainty introduced by the uncertain $B_0$ and $h$ parameter is ~4% for SSA, and ~20% for $g$. We will present a more detailed analysis of the model uncertainty in Section 6.4.

*6.3 Color filters*

Based on the analysis of Hapke modeling with clear filter data, we now fit the disk-resolved data taken through all color filters. Since the opposition surge cannot be constrained from our data, we decided to proceed with two cases for the modeling of color filter data. The first case has all five parameters free to fit, and the second case has $B_0$ set to 1.7 and $h$ set to 0.07, which are the averages of their respective best-fit values of all filters from the first case, in order to isolate the effect of uncertain opposition surge on the modeling of $g$-parameter and to directly study the wavelength dependence of the surface phase function. The best-fit parameters are listed in Table. 4 and Table. 5 and plotted in Fig. 9. The overall goodness plot of the best-fit Hapke model for F2 filter data is shown in Fig. 10 as an example. In this section, we will discuss the model results. In the next section (Section 6.4), we will provide a detailed analysis of the model uncertainties to back up our conclusions.

[Table. 4]

[Table. 5]

[Fig. 9]

[Fig. 10]

The model scatter for all filters and both cases are between 3-6% RMS, indicating satisfactory fits (Table. 4 and Table. 5, Fig. 9f, Fig. 10a). The model does not show any significant systematic trend along any scattering geometry parameters, $i$, $e$, and $\alpha$ (Fig. 10b). At high incidence angles ($i>60º$), corresponding to locations on the surface of Vesta near the terminator, the model scatter is slightly higher. Also the model scatter is slightly higher for moderate-to-high phase angles ($\alpha>50º$). This behavior indicates that the data with relatively worse fit are probably dominated by the southern mid-latitude areas (Fig. 2), presumably along the rim of the Rheasilvia basin, where the topographic slope is the highest on Vesta (Jaumann et al., 2012), the uncertainty in scattering geometry calculation is probably high, and the limitations of the photometric model are most prominent.  But overall, the Hapke model fits the surface well within $i<80°$ and $e<80°$.

Now we analyze the best-fit parameters, starting with roughness.  This parameter is probably the best-constrained parameter because it strongly affects the limb-darkening properties of the surface, which is only slightly affected by the SSA when the object is not very bright (SSA>0.9) but not any other parameters (McEwen, 1991).  Our $I/F$ data cover a range of phase angles from a few degrees to almost 90°, and nearly full range in the ($i$, $e$) space (Fig. 2).  The modeling of the roughness parameter should be unambiguous (Helfenstein et al., 1988).  Examining the model parameters, we noticed that the roughness parameters at all wavelengths for both cases concentrate within a small range of ±2° around ~18° (Table. 4 and Table. 5, Fig. 9c).  With the model uncertainty of ~9°, the scatters between the roughness parameters at seven wavelengths are not statistically significant.  The roughness parameter is presumably a geometric parameter that is only related to the mechanical properties of a surface.  On the other hand, since roughness measured from photometric techniques is dominated by the smallest topographic scales that cast shadows (Shepard and Campbell, 1998; Helfenstein and Shepard, 1999), strong inter-facet multiple scattering from a high-albedo surface can dilute the shadows and cause underestimate of the true topographic roughness.  Although the albedo of Vesta varies by >20% over the range of wavelengths of FC data, we do not see any wavelength dependence in the modeled roughness.  The lack of wavelength dependence indicates that either the multiple scattering on Vesta is not strong enough to cause significant change in the modeled photometric roughness, or the roughness on the surface of Vesta is caused by topographic shadows at scales much larger than the wavelengths of the FC data.  From the values fitted from all color filters, we found an average of 18°±4°.

Second we look at the opposition parameters. Setting both $B_0$ and $h$ parameters free does not improve the model RMS, again consistent with the opposition parameters not being well constrained by our disk-resolved data. The best-fit $B_0$ parameters are within a range of 1.4 and 1.9 (Table. 4, Fig. 9d). There is a slightly decreasing trend of $B_0$ with wavelength, but given the huge error bar for $B_0$ (Section 6.4), we cannot conclude that this trend is real or not. The best-fit $h$ parameters are within a range between 0.04 and 0.10. A slight trend of $h$ with wavelength is visible (Fig. 9e), and a correlation between $h$ and SSA is evident (Fig. 9h), too. However, due to the inter-coupling effects of the opposition parameters and the phase function parameter, $g$, and the large uncertainty of $h$ as well, one has to be extremely cautious in interpreting this trend. As discussed earlier, with the $B_0$ parameter of the SHOE consistently exceeding unity at all visible wavelengths, it is likely that CBOE also contributes to the opposition effect. The current understanding on CBOE is incomplete. Theoretical analysis suggests that the width of CBOE should be wavelength dependent (Hapke, 2002), although laboratory measurements (Nelson et al., 2000) and recent observations with the Lunar Reconnaissance Orbiter Camera (LROC) data of lunar surface (Hapke et al., 2012) did not show wavelength dependency for the width of CBOE. Our model based on the data with a minimum phase angle of ~5° cannot provide tests to the possibly present CBOE. Rather, we will only discuss the effects of uncertain opposition parameters on the modeling of other photometric parameters.

Next we consider the asymmetry factor, $g$, of the SSPF. Within the range of phase angles of our data, since both the SSPF and opposition effect affect the slope of the surface phase function, the determination of $g$ parameter strongly depends on the opposition parameters. Therefore, we begin our discussions with the case where the opposition parameters are fixed in the modeling. In this case, the variations in $g$-parameters with respect to wavelength reflect the variations in surface phase function, which is the combination of the wavelength dependency of both SSPF and opposition effect. Fig. 9b shows that the *g* values all concentrate in a small range between -0.220 and -0.231. Compared with the Minnaert *A* slope (Fig. 6), however, they show similarities on the overall spectral shapes – a dip with the lowest values (shallowest phase function) near 750 nm, corresponding to the highest albedo in Vesta's reflectance spectrum, and relatively steeper phase functions at the long- and short-wavelength ends in the wavelength range of our analysis. The wavelength dependencies of both Minnaert *A* slope and *g*-parameter are qualitatively consistent with the phase reddening behavior observed for Vesta from the ground and HEDs (Reddy et al., 2012b), where the spectrum of Vesta appears to be redder at higher phase angles in the range between 0º and 30º, and the 1-μm band appears to be deeper. We will present a detailed discussion on phase reddening of Vesta in Section 9.2.

For the case where $B_0$ and *h* are set free in our modeling, the *g*-parameter is relatively more backscattering between 550 nm and 850 nm, but appears to be slightly more forward scattering at 440 nm and near the center of the 1-μm band. Comparing with the spectra of $B_0$ and *h*, we suggest that such variations in *g* and the variations in the opposition parameters are correlated, where at 440 nm and within the 1-μm band, the opposition effect is strong, resulting in less backscattering SSPF. However, not being able to constrain the opposition parameters, we cannot draw conclusions about whether such a trend is real or not, nor should we go too far to interpret the possible trend.

The final modeled parameter is the SSA. This parameter provides an overall scaling factor to fit the model to data, and is only weakly affected by the other parameters for the available geometries of our data. It can therefore be well constrained. There is a good agreement between the two cases that we tested (Fig. 9a). The SSA spectrum shows similar features as compared to the reflectance spectrum of Vesta observed from the ground (e.g., McCord et al., 1970; Gaffey, 1997; Bus and Binzel, 2003), with a nearly linear spectral slope at wavelengths <650 nm, a maximum near 650-750 nm, and a broad absorption with a minimum near 950 nm (Fig. 9a).

Geometric albedo and Bond albedo are two derived parameters based on the best-fit parameters. Geometric albedo is essentially a model extrapolation of photometric data to zero phase angle. The extrapolation depends on the opposition effect, which can only be partially constrained by our data. The two cases that we experimented with resulted in geometric albedos differing by up to 9% (Fig. 9g). The uncertainty of geometric albedo is best estimated from disk-integrated phase function as shown in Figs. 1, 7, 8 where ground-based data covering phase angles of 2° are included. The uncertainty of geometric albedo should be dominated by the absolute radiometric calibration of our data and the limited data at small phase angle. We consider 10% as a reasonable estimate of the modeling uncertainty on geometric albedos in addition to the systematic radiometric calibration uncertainty. The shape of geometric albedo spectrum is similar to that of the SSA spectrum (Fig. 9a) and the spectrum of Vesta observed from the ground (e.g., Bus and Binzel, 2003) as well (Fig. 11). There appear to be discrepancies between the two cases that we tested in our modeling, and also discrepancies between the model geometric albedos and SMASS-II and HST observations (Li et al., 2011). Two causes are most probable. One is the uncertainties in the opposition parameters, and the other is the uncertainties in absolute radiometric calibration of FC data. The comparisons in Fig. 11 suggest that overall uncertainties of the modeled geometric albedos should be ~15%.

[Fig. 11]

The Bond albedo is the integration of the disk-integrated phase function weighted by sin($\alpha$), thereby usually dominated by the phase function within about 30º and 60º phase angles for the most common asteroidal phase functions. This range of phase angles is best observed by FC for Vesta. For this reason, the Bond albedos determined from our data (Table. 4 and Table. 5, Fig. 9h) are evidently reliable and independent of modeling process as suggested by the agreement of better than 1% between all methods in both cases of modeling. The uncertainty is only dominated by the absolute radiometric calibration of the FC, about 5% (Schröder et al., 2013a). Using the Bond albedo measurements through color filter and taking the solar flux through each filter (Schröder et al., 2013a) as the weighting factors, we derived a bolometric Bond albedo of 0.18 for Vesta, consistent with the measurement through the broadband clear filter, 0.19. This result is also consistent with the value of 0.15±0.03 for a composite V-type asteroid (Hicks et al., 2013). The uncertainty of the bolometric Bond albedo is about 5% (Schröder et al., 2013a), dominated by the radiometric calibration of the instrument.

*6.4 Model uncertainty analysis*

In this section we discuss the estimate of model uncertainties for the Hapke parameters that we derived. The absolute radiometric calibration uncertainty of ~5% (Schröder et al., 2013a) is a systematic uncertainty affecting all the reflectance data by the same factor at each wavelength, and is independent of, and in addition to, the uncertainty analysis we present here. However, the absolute radiometric uncertainty should only affect the various measurements of albedos, but not substantially affect the modeling of the surface phase function related parameters and the roughness parameter.

The estimate of uncertainties for Hapke parameters is complicated, mainly due to the imperfection of the model in describing the photometric behavior of planetary surfaces, and the complicated form of the model and the inter-coupled nature of the parameters violating the independency assumptions of parameters in the standard statistical analysis. The most common approach for error estimates of Hapke modeling is to define a 1-$\sigma$ error envelope in $\chi^2$ space where the $\chi^2$ reaches a certain value above the minimum, and search for the boundaries for one parameter while adjusting other parameters accordingly to compensate for the inter-coupling of parameters. We define our error envelope at twice the minimum $\chi^2$. This method is similar to the method proposed by Helfenstein and Shepard (2011), but does not use the distribution of model residuals to define the error envelope.

Another difficulty in estimating the errors for our model parameters is that the opposition parameters, especially $B_0$, cannot be well constrained by our data. We found that even if $B_0$ is set to 10, a large value that is unphysical, we can still adjust other parameters accordingly so that the $\chi^2$ only increases by less than 10%. In this case, the $\chi^2$ envelope does not provide any constraint to estimate the uncertainty for $B_0$. Therefore we had to use other constraints, in particular, the geometric albedo because it is extremely sensitive to $B_0$. The geometric albedo usually cannot be directly measured for main belt asteroids because the geometry is not accessible from the ground. But some empirical models, such as the IAU H-G phase function model, usually provide good estimates of geometric albedo by extrapolating observations to zero phase angle. We assume that the true geometric albedos of Vesta through FC filters are within ±20% of the modeled values. With the assumption of ±20% for the modeled geometric albedos, we can put constraints on the opposition parameters and other parameters as well, if needed.

The error estimate for $B_0$ is illustrated in Fig. 12 for F2 filter.  The constraints for both the lower and upper boundaries of $B_0$ are provided by the geometric albedo constraint (Fig. 12b). While the error envelope does provide a lower limit of 0.1, which is not particularly useful, it cannot provide any constraint even up to a value of 10.  The phase functions with $B_0$ between 0 and 3 are plotted in Fig. 12c, showing that the largest difference between these phase functions is within 15º phase angle, while the effect of opposition is negligible outside of this region.  It demonstrates that geometric albedo does provide a reasonably good constraint on $B_0$.  This method yielded a range of $B_0$ as $1.1 - 2.3$, or an error bar of ±0.6 for the average value of 1.7 over all FC wavelengths.  Similar to $B_0$, the range of the width parameter, $h$, can only be found from the geometric albedo constraint to be $0.03 - 0.15$, or an error bar of +0.08/-0.04 for the wavelength average value of 0.07.

[Fig. 12]

For the phase function parameter, $g$, the situation is more complicated.  Using the ranges of $B_0$ and $h$ determined above, one boundary of $g$ can be determined by the $\chi^2$ error envelope to be -0.33.  However, when $g$ approaches this value, both $B_0$ and $h$ land on their respective boundaries, implying that this boundary of $g$ is in fact indirectly determined by the geometric albedo constraint as well.  The other boundary of $g$ is directly determined by the geometric albedo constraint as -0.17.  The error bar for $g$ is therefore +0.07/-0.09.

A similar situation occurs for SSA.  Both boundaries of the SSA are set by the $\chi^2$ error envelope, but the opposition parameters hit their boundaries derived from the geometric albedo constraints.  The range of SSA is $0.41 - 0.57$, or ±0.08, or 16%.

The roughness parameter is the only parameter that does not strongly depend on the ranges of other parameters to set its model uncertainties, partly owing to the fact that it is mostly determined by limb-darkening characteristics. The $\chi^2$ error envelope results in an error bar of ±9°. Although $B_0$ and $h$ still hit their respective extreme values, we tested the error estimate with subsets of reflectance data, e.g., between 40° and 50° phase angles, and found a similar $\chi^2$ structure. In addition, even if we relaxed the ranges of $B_0$, $h$, and $g$ in the analysis, the error estimate for roughness does not change. In the final test, we fixed $B_0$, $h$, and $g$ at their best-fit value, adjusted only the SSA, and retrieved similar results for the error bar of roughness.

The error bars that we discussed in this section are specifically for the F2 filter data at 554 nm wavelength. Analyses for other filters show similar results as expected because of the comparable quality of data through all filters and their similar geometries. The error bars of $B_0$, $h$, $g$, and roughness are nearly identical, and the relative error bars for the SSA given as a percentage remain similar.

We would like to point out that the model uncertainties that we derived in this section should be considered "systematic" as introduced by the limitation of our data in the scattering geometries, the statistic scatters in the modeling process, and possibly the imperfection of the Hapke model in describing the photometric behavior of planetary surfaces. They are "systematic" in the sense that if, for some reason, the best-fit value for one parameter is off by, e.g., 20% in one filter, then the same 20% applies to the results from all filters. If we want to compare the results of Vesta with other objects, then these error bars, as well as the absolute calibration errors must be considered. On the other hand, such error bars should not affect the comparisons between different filters, and do not affect the overall shape of the spectra of those parameters.

Rather, the relative uncertainties from filter to filter should be characterized by the scatter of data points in Fig. 9 only, which are typically a few percent at most. The overall wavelength trend of those parameters should be more reliable than their absolute values.

## 7. Comparisons between various photometric models

One goal of this study is to determine what model best describes the photometric behavior of Vesta's surface. For this purpose, we present comparisons of various photometric models to the surface of Vesta in this section.

Generally, all photometric models contain two components, including the "disk-function" or "limb-darkening function" that describes the dependence of reflectance on local scattering geometry ($i$ and $e$) at a particular phase angle, and the phase function that describes the dependence of reflectance on phase angle. Hapke's model includes both components. But many other empirical models, such as the Minnaert model we studied in Section 5, the parameterless Akimov model (Shkuratov et al., 2011) and Lommel-Seeliger-Lambertian (LS-Lambertian) model (McEwen, 1991, 1996) as applied to Vesta by Schröder et al. (2013b), and the pure Lommel-Seeliger (LS) model, provide only the disk function component, while the modeling of phase function is realized by an analytically simple, empirical function, such as a polynomial, to the reflectance that has been corrected for local scattering geometry (e.g., Hillier et al., 1998; Schröder et al., 2013b). Therefore, in our comparisons discussed in this section, we will focus on the performance of these models in the disk-function component only. In addition, the disk-function provides corrections to local scattering geometry before we can model the phase function, and therefore usually dominates the overall model quality.

The surface of Vesta shows large photometric and color heterogeneities (Reddy et al., 2012a; Schröder et al., 2013b) and complicated topography (Jaumann et al., 2012). We have to look at the average surface of Vesta to study its global photometric behaviors and compare with various models instead of using the scans from any single images. Starting from the binned $I/F$ data from RC3 and RC3b images as described earlier in our photometric modeling, we chose those within 2º along the apparent photometric equator (where $|i{\pm}e|{=}\alpha$) and the mirror meridian (where $i{=}e$) at two phase angles, 37º and 8º, respectively, as plotted in Fig. 13. Then we calculated the predicted reflectance along the corresponding photometric equators and mirror meridians from all five models mentioned above, and compared them with the measurements.

The simple LS model, which has shown to be able to describe the dark surfaces of primitive type asteroids and comets and that of the Moon, generates a flat reflectance distribution along the mirror meridian regardless of phase angle, inconsistent with that of Vesta at relatively high phase angles. Note that at low phase angles, the surface of Vesta is relatively flat within $e{<}60$º. In addition, the LS model predicts strong limb-brightening at $e{>}70$º along photometric equator, inconsistent with the data. The LS-Lambertian model adds limb-darkening to the LS model with the Lambertian term, representing an empirical improvement to the LS model. The best-fit LS-Lambertian model for Vesta (Schröder et al., 2013b) is consistent with the reflectance of Vesta along the photometric equators at phase angles $<{\sim}75$º at 37º phase angle, and $<{\sim}50$º at 8º phase angle. But along the mirror meridians, the LS-Lambertian model is inconsistent with observations at both phase angles we examined. The Minnaert model generally fits the surface of Vesta at $e{<}60$º. But similar to the LS and LS-Lambertian model, it also suffers strong limb-brightening at e>60º along the photometric equator. Both the Hapke model and the Akimov model were able to achieve satisfactory fit to all four cases we tested here. Along the photometric equator, the Akimov model shows a slightly better fit to the data than the Hapke model, which underestimates the reflectance by a few percent at low to moderate emission angles.  But along the mirror meridian, the Hapke model delivers a slightly better fit within the range of data we used in the fit, $e<80º$, and the Akimov model overestimates the reflectance near limb at $e>60º$.

[Fig. 13]

The inconsistency between the photometric behavior of Vesta and the LS model is probably due to the moderate multiple scattering on Vesta caused by its relatively high albedo among rocky asteroids, because the LS model assumes that single scattering dominates the total reflectance.  The LS-Lambertian model does improve the fit by adding the empirical, multiple scattering dominant term.  But for the case of Vesta, the fit is still not adequately good, probably because the combination of single scattering and multiple scattering on Vesta is more complicated than the simple linear combination for the two components.

Our results are consistent with those reported by Schröder et al. (2013b), although their studies were focused on the broadband filter.  After comparing the Akimov model, Minnaert model, and LS model, they identified that the Akimov model to be the most satisfactory of the three for Vesta, and the LS model is the least.  Our comparisons show that the Hapke model we derived delivers comparable quality to describe the photometric behaviors of Vesta as the Akimov model.

In summary, the light scattering properties of Vesta are best described by the Hapke model and Akimov model.  Minnaert model can produce satisfactory fit within ~60º emission angles.

For the purpose of photometric corrections, we recommend the Hapke model and the Akimov model.

## 8. Photometric variations

Dawn FC images and VIR spectra confirmed that the significant albedo and spectral variations across the surface of Vesta are caused by compositional heterogeneity on the surface (Reddy et al., 2012a; De Sanctis et al., 2012; 2013). Schröder et al. (2013b) uncovered the variations of phase function and/or roughness over the surface from FC images. We performed a detailed study of the albedo variations based on the photometric models we derived. The detailed comparisons between Dawn albedo maps and previous HST and ground-based observations are discussed in a separate paper by Reddy et al. (2013).

For this study, we generated photometrically corrected mosaics using the best-fit Hapke parameters with $B_0$=1.7 and $h$=0.07 fixed (Table. 5). We do not expect the opposition parameters to affect the results of the photometric correction because almost all images we used were outside of the opposition surge. Fig. 14 shows the mosaics derived from RC3b (upper panel) and RC3 (lower panel) through F3 filter (749 nm), at average phase angles of ~10º and ~37º, respectively. The mosaics are photometrically corrected radiance factor at $i$=30º, $e$=0º, and $\alpha$=30º, and projected in a sinusoidal projection that preserves surface area. The RC3b mosaic provides more coverage in the southern hemisphere due to the more southern sub-spacecraft latitude (~-20º) than the RC3 mosaic (~7º). Compared with each other, the mosaic derived from RC3b images appears to be cleaner than the one derived from the RC3 images, with almost no bright artifacts of up to a few km in size. These artifacts in the RC3 mosaic are presumably introduced by photometric correction to extreme scattering angles and possibly shadows within craters at higher phase angles of RC3 than RC3b. Therefore, we consider that the RC3b mosaic has overall higher quality than the RC3 mosaic. The brightness distribution in our reflectance mosaic is consistent with that of Schröder et al. (2013b), who used the Akimov model to photometrically correct the RC3b images.

[Fig. 14]

The albedo mosaic of Vesta shows predominant albedo variations across the mapped area. The albedo variations are both qualitatively and quantitatively consistent with earlier ground-based and HST observations, albeit at much lower resolutions (Reddy et al., 2013). The relatively dark areas mostly concentrate between 60º and 190º longitude (all longitudes are in the Claudia coordinate system; see Russell et al., 2012, for the definition of the this coordinate system and Reddy et al., 2013, for a comparison of this coordinate system with others) and north of -30º, while the relatively bright areas are distributed outside of this region (Fig. 15). The Rheasilvia basin appears to be systematically brighter than other areas on Vesta, and is ~10% brighter than the global average and ~18% brighter than the northern hemisphere (Fig. 15a).

[Fig. 15]

There are many small, localized bright and dark areas <50 km in sizes with clear boundaries from the surrounding areas. The darkest area as observed at a spatial sampling of ~0.5 km/pix is Aricia Tholus, located at ~11º latitude and ~161º longitude, and ~100 km to the western rim of Marcia crater, with a reflectance of 0.08-0.10. There are several bright areas with reflectance of 0.23-0.25 located in the ejecta of large craters, including Canuleia, Justina, and Tuccia, located at latitudes between -35º and -40º. A small (10×2 km in size) but extremely bright strip located on the wall of an unnamed crater centered at -64º latitude and 358º longitude has a reflectance of >2× of the average reflectance. Its reflectance cannot be accurately compared with other areas due to its small size and its location with a high incidence angle posing difficulties for photometric correction, but this area might be the brightest on Vesta.

The properties and origins of these bright and dark areas are the subject of intensive studies (e.g., McCord et al., 2012; Li et al., 2012; and references therein). Preliminary results suggest that dark areas might be exogenous carbonaceous material delivered to Vesta through low-speed impacts (McCord et al., 2012; Reddy et al., 2012c), and bright areas are the exposures of materials originally formed on Vesta with the least amount of mixing with exogenous materials or alterations (Li et al., 2012). Regolith mixing processes induced by impacts at all scales are responsible for the overall appearance of the entire surface of Vesta (Pieters et al., 2012). This scenario is completely different from that on the Moon, where mineralogical variations dominate the albedo variations.

Fig. 16 shows the histograms of surface reflectance of Vesta produced from the mosaics shown in Fig. 14. The histogram derived from RC3 mosaic has a slightly wider base than that from RC3b mosaic, presumably due to more prominent artifacts at higher phase angles as discussed above. The overall shape of the histogram is single-peaked, although a small shoulder appears on the low-reflectance side of RC3b histogram but not the RC3 histogram. Whether the shoulder is real or not is uncertain due to the incomplete surface coverage of our mosaics and the imperfection of photometric correction. But the reflectance distribution of Vesta's surface is certainly different from that of the Moon, which has a clear bimodal distribution from the dark mare and the bright highlands (e.g., Helfenstein and Veverka, 1987; Hillier et al., 1999; Yokota et al., 2011).  The single-peaked reflectance distribution on Vesta is consistent with the scenario that its albedo is dominated by regolith mixing processes at both global and local scales rather than compositional variations.  The average reflectance of the surface of Vesta covered in the RC3b mosaic is 0.18 at 554 nm, and the full-width-at-half-max (FWHM) of the reflectance distribution is about 17% of the average, consistent with previous HST observations (Li et al., 2010).  The FWHM of Vesta's global reflectance distribution is much narrower than the range of reflectance seen at ~0.5 km resolution, also consistent with significant regolith mixing at the global scale.

[Fig. 16]

The bolometric Bond albedo map of Vesta based on RC3b mosaics is shown in Fig. 17, produced by scaling the photometrically corrected mosaics of each color filter by the corresponding Bond albedos (Table. 4 and Table. 5), then taking the average with the corresponding solar flux through each filter as the weighting factors.  The map is very similar to the reflectance mosaics shown in Fig. 14, suggesting albedo variations rather than other photometric variations (in phase function and/or roughness etc.) dominate the bolometric Bond albedo variations on Vesta.

[Fig. 17]

## 9.  Discussion

*9.1 Multiple scattering*

The traditional method of correcting imaging and spectroscopic data to a common illumination and viewing geometry is to calculate the model predicted images from the shape model under the same geometries of the corresponding data, assuming uniform photometric properties over the surface. The ratio of the images and the corresponding model prediction is taken as the reflectance deviation from a uniform surface. The ratios are then scaled to a standard geometry, usually $i$=30º and $e$=0º, to compare images taken at various geometries and with laboratory measurements of samples. This approach implicitly assumes that reflectance is proportional to albedo (SSA or normal albedo). This assumption is true only when multiple scattering, which is non-linear to albedo, is a low or negligible component of the total reflectance, such as the surfaces of dark asteroids and cometary nuclei. Vesta is among the brightest asteroids in the main asteroid belt, although still much darker than icy bodies. We have to verify the linearity in order to validate the traditional approach for photometric corrections.

Based on Hapke model, we calculated the predicted $I/F$ as a function of SSA (Fig. 18a), assuming other photometric parameters as those modeled from F2 filter (554 nm) for Vesta. Within ±30% of the best-fit SSA for this filter, a linear relationship is a good approximation for Vesta. However, for the brightest area on Vesta, where the reflectance is probably >2× the global average, the proportionality does not hold. Therefore, based on the reflectance histogram of Vesta (0), we conclude that for Vesta, the traditional approach of photometric correction should be a good approximation for >99% of the surface of Vesta. One has to be cautious, though, when considering localized areas with reflectance higher than >50% of the global average. In addition, Schröder et al. (2013b) demonstrated that there exist variations in the surface phase function, possibly caused by variations in the SSPF and the roughness from crater walls to floors. It is uncertain whether such variations on photometric variations other than albedo are the cause of the weak seams shown in the photometrically corrected mosaics. This is a subject of future work.

How does the ~5% absolute radiometric calibration uncertainty and the possible residual stray light affect the photometric modeling?  Based on the linear relationship between reflectance and albedo within ±30% of the average, we consider that the radiometric calibration uncertainty only affects the measurement and modeling of albedo, but has negligible effects on other photometric parameters that we derived.   In the development of this work, we performed modeling with several versions of the radiometric calibration that differed by up to 10%.  The albedos were the most affected model parameters.  Other parameters, such as phase function parameters, including $g$, $B_0$, and $h$, only changed within the scatter among different wavelengths; and the change of roughness parameter was only 1º.  Therefore our modeled parameters, except for albedo, should not substantially change if any future improvements to the radiometric calibration produce values that are within the current uncertainties.

For Vesta, the fraction of multiple scattering is calculated to be 15-30% within 80º phase angle based on our best-fit Hapke model parameters (Fig. 18b).  This is a non-negligible fraction.  But as demonstrated in the previous paragraph, this fraction does not significantly affect the linear dependence of reflectance on albedo.  In our Hapke modeling process, we experimented with the anisotropic multiple scattering model (Hapke, 2002).  The model does not show any improvement for our data in terms of returning smaller model residuals.  We also experimented with the most recent improvement of the Hapke's model with the porosity parameter (Hapke, 2008).  With most initial parameter sets that we tried, the best-fit filling factor is zero, and the model degenerates to the version without porosity included.  The effects of multiple scattering are not appreciable in our experiments for the case of Vesta.  The difference between using the isotropic multiple scattering model and anisotropic model may just be comparable to the scatter in our data, due to the large albedo variations on Vesta.  Therefore our conclusion is that for Vesta, which has ~0.50 SSA, multiple scattering does not have appreciable effect on the photometric modeling under Hapke's theory.  The effect of multiple scattering might be appreciable only when the SSA reaches ~90%, at which point multiple scattering has comparable contributions to the total reflectance as single scattering.  Of course, the Hapke theory itself contains a number of assumptions, and it is not possible for us to test and validate all model assumptions based on our available data.

*9.2 Phase reddening*

Asteroids often show redder spectral slopes when observed at higher phase angles (Bowell and Lumme, 1979; Clark et al., 2002; Kitazato et al., 2008), the so-called phase reddening. Laboratory bidirectional reflectance measurements of powdered minerals and asteroid analog materials show similar reddening effect (Gradie et al., 1980; Gradie and Veverka, 1982).  Clark et al. (2002) quantitatively studied the phase reddening behaviors of Eros based on their photometric modeling of this asteroid derived from the Near-Earth Asteroid Rendezvous (NEAR) Near-Infrared Spectrometer (NIS) data at phase angles from 0º to ~110º and wavelengths from 0.8 μm to 2.4 μm.  Reddy et al. (2012b) performed a detailed study of phase reddening of Vesta from ground-based, multispectral observations, and laboratory measurements of HED samples. In this section, we will perform a quantitative analysis of the disk-integrated phase reddening behavior of Vesta based on our photometric models, and compare with Eros and ground-based observations of Vesta and laboratory measurements of HEDs.

To study the difference in the phase functions of Vesta at various wavelengths, we calculated the disk-integrated phase functions based on the best-fit Minnaert and Hapke models we discussed in Sections 5 and 6.3, respectively, and calculated the ratio of phase functions between all color filters and F3 at 748 nm. For Hapke models, to avoid the potential uncertainties introduced by fitting the highly unconstrained $B_0$ and $h$, we used the models with $B_0$ and $h$ fixed. The phase functions calculated from Hapke models from 0º to 120º phase angles are shown in Fig. 19a, and the ratios are shown in Fig. 19b and Fig. 19c, where Fig. 19c is the zoom in of 0º to 30º phase angles to compare with ground-based observations of Vesta. The phase function ratios calculated from Minnaert models are similar to those calculated from Hapke model as shown in Fig. 19 at phase angles <60º. At higher phase angles, the ratios predicted by Minnaert models are lower than those predicted by Hapke models by up to 10%. We will focus our discussion based on the Hapke model predictions of phase function ratios for two reasons: 1) the previous ground-based observational data for Vesta are only available at phase angles <30º, at which range both Hapke models and Minnaert models are similar to each other; and 2) similar studies for Eros (Clark et al., 2002) were based on the Hapke models.

[Fig. 19]

The phase function ratios (Fig. 19) suggest steeper phase functions of Vesta at all other wavelengths compared to 748 nm. At wavelengths shorter than 748 nm, the steeper phase functions at 653 nm, 554 nm, and 438 nm are consistent with a reddening of the spectrum of Vesta with increasing phase angle within this wavelength region as previously observed (Reddy et al., 2012b). However, the numerical relationship between spectral slope and phase angles provided by Reddy et al. yielded values of ~0.97, ~0.94, and ~0.88 at 25º phase angle for the ratios of phase functions at those three wavelengths, significantly lower than the values of ~0.99, ~0.98, and ~0.96, respectively, derived from our photometric models. Our models, based on the observations of Vesta from Dawn FC, predicted less phase reddening for Vesta than that observed from the ground. The phase functions at three wavelengths, 829 nm, 917 nm, and 965 nm are within the 1-μm band, with the 917 nm one closest to the band center of howardites, the dominant compositional component on the surface of Vesta (De Sanctis et al., 2012). The steeper phase functions at those three wavelengths are consistent with deeper bands at higher phase angles. However, quantitatively, the numerical relationship for Band-I depth with phase angle provided by Reddy et al. predicted a phase ratio of ~0.85 between 0.917 nm and 0.748 nm at a phase angle of 25º, compared to our value of ~0.95, suggesting a much higher phase reddening effect observed from the ground than by the FC. Reddy et al. also reported decreasing band depths with phase angle at phase angles higher than 40º-60º for eucrite and terrestrial low-calcium pyroxene samples they measured. Our models (Fig. 19b) showed that for the case of Vesta, the phase reddening could be weakening at phase angles >80º.

Compared to Eros, Vesta showed similar or perhaps slightly stronger phase reddening. For Eros, as reported by Clark et al. (2002), the ratio of the phase functions outside of the 1- and 2-μm bands (2.320 μm/1.486 μm) suggested a shallower slope at longer wavelengths, with about 0.07%/deg within 100º phase angle. The ratios of the phase functions outside the absorption band and near the band center for both 1-μm (1.486 μm/0.946 μm) band and 2-μm band (2.320 μm/1.932 μm) are consistent with deeper bands at higher phase angles within 100º phase angle. The ratio of the phase functions at the band centers (1.932 μm/0.946 μm) is consistent with a stronger band depth increase for the 1-μm band with phase angle than that for the 2-μm band, qualitatively similar to the case of Vesta as observed from the ground (Reddy et al., 2012b). Note that the phase function ratios we studied are reciprocal to those calculated by Clark et al. for Eros. For Vesta, the phase function ratios are slightly higher than those for Eros (ignoring phase angles <10º that are possibly within the opposition surge), with similar shapes with respect to phase angle.

The exact reasons for phase reddening are unknown, but recent studies suggest that it is likely caused by increasing contribution of multiple scattering at higher phase angles as the albedo increases (e.g., Hapke et al., 2012). This interpretation is generally consistent with the phase reddening of Vesta. Fig. 9b shows that the $g$ parameter, which for the case of $B_0$ and $h$ fixed describes the combined effect of SSPF and SHOE, is likely independent of wavelengths, indicating that phase reddening is not likely a single-scattering phenomenon. The 15-30% multiple scattering from 10º to 70º phase angle is certainly a considerable fraction of the total reflectance (Fig. 18), supporting the multiple-scattering interpretation of phase reddening on Vesta. Compared to other asteroids, the albedo of Vesta is high, and is about 40% higher than that of Eros (see next section and Table 6). Therefore, we expect higher contribution of multiple scattering on Vesta than on Eros, and in turn stronger phase reddening, which is consistent with our early comparisons using spacecraft observations and ground-based observations.

*9.3 Comparisons with other solar system objects*

Compared to other asteroids and comets with their photometric properties modeled from spacecraft images with the Hapke model (Table. 6), Vesta has a high albedo. The cometary nuclei all have geometric albedos about 4-6%, and Bond albedo of ~1%, and Vesta has a geometric albedo of 36%, and a Bond albedo of 18%. The E-type asteroid Steins has a slightly higher SSA than Vesta. But due to the less prominent opposition surge of Steins, its geometric albedo is comparable or slightly lower than Vesta. The phase function of Vesta and the opposition surge are similar to those of S-type asteroids and lunar maria, which have an asymmetry factor of the SSPF between -0.35 and -0.21, as also noticed by Belskaya and Shevchenko (2000). Dark, carbonaceous-type asteroids have steeper phase functions than Vesta with asymmetry factors between -0.4 and -0.5, with Mathilde being a notable exception. The roughness parameter of all those asteroids and cometary nuclei are between 20º and 30º, except for Ceres, which has a value of 44º and was interpreted as possibly different surface regolith structure or due to lack of data at sufficiently high phase angle for a reliable determination of the roughness parameter (Li et al., 2006). We cannot make reliable comparisons for the opposition parameters due to the small number of reliable modeling from limited data at low phase angles. But in general, the opposition amplitudes of almost all asteroids listed in Table. 6 are between 1.0 and 2.0, and the width parameters range from 0.02 to 0.08, suggesting the possible existence of CBOE on all of them with an angular width of 2º-10º in phase angle. In short, Vesta's global photometric properties are similar to those of S-type asteroids, but with an albedo about twice as high.

## 10. Summary

We performed a detailed photometric analysis for Vesta using the Dawn FC data obtained during and prior to the Survey Orbit at an altitude of ~3000 km with pixel sizes >0.5 km/pix through seven narrowband color filters from 438 nm to 961 nm and one broadband clear filter centered at ~700 nm. The disk-integrated phase function constructed from approach data through the clear filter when Vesta is smaller than the FOV covers phase angles ~23º to 108º. The phase function of Vesta can be best described by an empirical H-G phase function model with H=3.2 and G=0.28 at phase angles <80º. At higher phase angles, Vesta's brightness is higher than the model predicted. This discrepancy should be due to the limitation of the H-G model.  The geometric albedo of Vesta based on our phase function model is 0.36, consistent with earlier measurements.

We extracted disk-resolved radiance factor measurements from the disk-resolved data at a spatial resolution of ~1 km/pix for all seven color filters and the broadband filter, and used the high-resolution topography model of Vesta at a resolution of 80 m/pix as derived from Dawn data to calculate the local scattering geometry.  We applied both the Minnaert model and a five-parameter version of Hapke model to the disk-resolved data with incidence angles and emission angles both <80º.  The results are summarized here:

1.  The modeled Minnaert $k$ parameters have similar values at all wavelengths, and show a linear relationship with phase angle.  At zero phase angle, the extrapolated values of $k$ is ~0.54, with no obvious trend with wavelength, representing a flat disk of Vesta at low phase angles, similar to that of the Moon and cometary nuclei.  The slope of Minnaert $k$ with respect to phase angle is ~0.0037/deg, with no obvious dependence observed.

2.  The phase slope of Vesta as represented by the phase slope of the Minnaert albedos parameter shows a slight correlation with wavelength, where within the 1-μm band center, the phase function is steeper, consistent with the phase reddening of Vesta observed from the ground.

3.  The best-fit Hapke parameters of Vesta through FC color filters are: $\bar{\theta}$ is from 16º to 20º; $g$ is within a small range of -0.22 to -0.23; $B_0$ is ~1.7; and $h$ is ~0.07.  These parameters show weak or no dependence on wavelength.  The SSA is 0.50 at 554 nm, 0.54 at 748 nm, and displays a similar spectrum as previously observed reflectance spectrum of Vesta. The bolometric Bond albedo of Vesta is 0.18.

4. We estimated the Hapke model uncertainties based on two criteria: 1) the $\chi^2$ reaches twice the minimum; and 2) the geometric albedo of Vesta has been determined to within ±20% accuracy. The uncertainty of $B_0$ is ±0.6. The uncertain range of $h$ is 0.03 to 0.15. The uncertainty of $g$ is ±0.08. The uncertainty of the SSA is ±16%. And the uncertainty of the roughness parameter is ±9º. We estimated that the overall uncertainties of geometric albedos should be ~15%, and the uncertainties of Bond albedos should be comparable to the radiometric calibration uncertainties of the camera, including stray light removal, ~5% total.

Based on our photometric modeling, below is a summary of the photometric properties of Vesta's surface.

1. The overall photometric properties of Vesta is best described by the Hapke model and the Akimov disk function, which have comparable model accuracy and slightly different performance at different scattering angles. The Minnaert model suffers from predicting strong limb-brightening that is not observed on Vesta. The LS model and LS-Lambertian model cannot fit the reflectance along the photometric equator, and produce strong limb-brightening.

2. The average radiance factor of Vesta at 554 nm is 0.18 (corrected to $i$=30º, $e$=0º), with a FWHM of 17% of the average at a spatial sampling of 0.5 km/pix. The histogram of the reflectance distribution is single-peaked, clearly different from that of the Moon, and probably a consequence of significant regolith mixing at all spatial scales.

3.  The albedo of Vesta shows obvious global scale variations, where the relatively dark areas are concentrated between 60º and 190º in longitude and north of -30º in latitude. The Rheasilvia basin is about ~10% brighter than the global average, and ~18% brighter than the northern hemisphere. The darkest area on Vesta observed from Dawn FC at 0.5 km/pix is Aricia Tholus (11º latitude and ~161º longitude), with a radiance factor of 0.08-0.10. The brightest areas on Vesta is on the wall of an unnamed crater centered at -64º latitude and 358º longitude, with a reflectance of at least 2× of global average.

4.  Vesta clearly shows phase reddening effect. But we measured weaker phase reddening from Dawn FC data than that reported from ground-based observations at comparable phase angles, and from laboratory measurement of HEDs. The phase reddening of Vesta measured from Dawn FC is comparable to or slightly stronger than that of Eros previously measured by NEAR NIS.

5.  Compared to other asteroids imaged by spacecraft from close distances, Vesta's photometric properties are similar to those of S-type asteroids, but with an albedo significantly higher. Vesta is among the highest of all asteroids.

Based on our work presented in this paper, we recommend the Hapke model or the Akimov model for the first order photometric corrections to the FC data of Vesta. The traditional approach for photometric correction is validated for >99% of the surface of Vesta where the albedo is within ±30% of the global average.

**Acknowledgements**

This work is supported by NASA's Dawn at Vesta Participating Scientist Program and NASA's Dawn Mission through the Discovery Program. JYL is supported by NASA's Dawn at Vesta Participating Scientist Program through Grants NNX10AR56G to University of Maryland at College Park and NNX13AB82G to Planetary Science Institute. Part of this work was carried out at the Jet Propulsion Laboratory, California Institute of Technology, under a contract with NASA. The Framing Camera project is financially supported by the Max Planck Society and the German Space Agency, DLR. The authors are extremely grateful to the wonderful engineering, operations, and instrument teams who made the Vesta phase of Dawn mission a great success. We thank Dr. Beth Clark and Dr. Paul Helfenstein for their careful and rigorous reviews, which helped improve the manuscript significantly.

Table. 1. Sequence of images used in our study. We used VSA_ON_01 to VSA_ON_18 data to study the disk-integrated phase function, and VSA_ON_19 through all Survey data (shaded rows) in disk-resolved analysis.

| Sequence Name | Start Time (UTC) | Duration (hrs) | Filter | Pixel Size (km/pix) | Phase Angle (deg) | Sub-S/C Latitude (deg) | Sub-Solar Latitude (deg) |
|---|---|---|---|---|---|---|---|
| VSA_ON_01 | 2011-05-03T13:36 | 0.48 | F1 | 114 | 42.7 | -9.4 | -21.3 |
| VSA_ON_02 | 2011-05-10T07:03 | 1.42 | F1 | 94.6 | 42.5 | -10.6 | -21.9 |
| VSA_ON_03 | 2011-05-17T12:57 | 0.48 | F1 | 75.8 | 41.9 | -12.1 | -22.6 |
| VSA_ON_04 | 2011-05-24T08:52 | 0.48 | F1 | 60.4 | 40.9 | -13.7 | -23.2 |
| VSA_ON_05 | 2011-06-01T06:37 | 0.48 | F1 | 45.3 | 39.5 | -15.8 | -23.8 |
| VSA_ON_06 | 2011-06-08T15:24 | 1.43 | F1 | 33.0 | 37.2 | -18.3 | -24.3 |
| VSA_ON_07 | 2011-06-14T13:38 | 0.95 | F1 | 24.8 | 34.8 | -20.9 | -24.7 |
| VSA_ON_08 | 2011-06-17T12:38 | 0.95 | F1 | 21.2 | 33.4 | -22.4 | -24.9 |
| VSA_ON_09 | 2011-06-20T13:38 | 0.95 | F1 | 17.9 | 31.8 | -24.1 | -25.1 |
| VSA_ON_10 | 2011-06-24T04:08 | 0.95 | F1 | 14.3 | 29.8 | -25.5 | -25.3 |
| VSA_ON_12 (RC1) | 2011-06-30T04:55 | 5.22 | All | 9.2 | 26.3 | -32.1 | -25.7 |
| VSA_ON_13 | 2011-07-04T00:40 | 1.90 | F1 | 6.6 | 23.6 | -38.0 | -25.9 |
| VSA_ON_15 (RC2) | 2011-07-09T23:52 | 5.22 | All | 3.4 | 29.3 | -53.9 | -26.2 |
| VSA_ON_16 | 2011-07-13T03:10 | 1.90 | F1 | 2.3 | 42.7 | -69.2 | -26.3 |
| VSA_ON_17 | 2011-07-17T03:39 | 1.90 | F1 | 1.3 | 80.7 | -69.3 | -26.5 |
| VSA_ON_18 | 2011-07-18T20:40 | 1.90 | F1 | 1.0 | 107.5 | -43.5 | -26.6 |
| VSA_ON_19 | 2011-07-23T18:16 | 2.95 | F1 | 0.52 | 61.4 - 67.7 | +34.0 to +40.3 | -26.7 |
| VSA_ON_20 (RC3) | 2011-07-24T06:00 | 5.27 | All | 0.51 | 32.0 - 43.0 | +4.0 to +15.2 | -26.7 |
| VSA_ON_20 (RC3b) | 2011-07-24T20:36 | 5.27 | All | 0.52 | 7.7 - 13.6 | -16 to -27 | -26.8 |
| VSA_ON_21 | 2011-07-26T03:33 | 1.23 | F1 | 0.53 | 53.6 - 56.0 | -80.4 to -82.8 | -26.8 |
| VSA_ON_22 | 2011-07-31T11:31 | 1.17 | F1 | 0.38 | 25.5 - 29.2 | -2.6 to +1.1 | -27.0 |
| VSA_ON_23 | 2011-08-06T02:44 | 1.16 | F1 | 0.28 | 61.7 - 67.7 | +34.0 to +40.1 | -27.1 |
| VSA_ON_23 (C0) | 2011-08-06T07:20 | 5.58 | F1 | 0.28 | 16.9 - 44.2 | -13.2 to +16.0 | -27.1 |
| Survey | 2011-08-11T18:13 | - | All | 0.28 | 11.0 - 80.0 | Varies | -27.3 |

**Table. 2.**  Hapke model fit to disk-resolved data of Vesta from Approach and Survey taken through clear filter.

|  | SSA | $B_0$ | $h$ | $g$ | $\theta$ (deg) | RMS (%) | $A_{geo}$ | $A_{Bond}$ | Plot color |
|---|---|---|---|---|---|---|---|---|---|
| Case 1: | 0.511 | (1.03) | (0.04) | -0.294 | 17.5 | 5.1 | 0.365 | 0.190 | Red |
| Case 2: | 0.501 | (1.03) | 0.076 | -0.271 | 17.8 | 5.0 | 0.332 | 0.190 | Green |
| Case 3: | 0.491 | 1.66 | 0.076 | -0.233 | 18.3 | 4.8 | 0.369 | 0.191 | Blue |

**Table. 3.** Hapke model fit to disk-integrated phase function of Vesta from Approach data taken through clear filter.

| | SSA | $B_0$ | $h$ | $g$ | RMS (%) | $A_{geo}$ | $A_{Bond}$ | Plot color |
|---|---|---|---|---|---|---|---|---|
| Case 1: | 0.491 | (1.03) | (0.04) | -0.280 | 5.0 | 0.333 | 0.177 | Red |
| Case 2: | 0.487 | (1.00) | 0.054 | -0.272 | 5.0 | 0.318 | 0.177 | Orange |
| Case 3: | 0.477 | 1.77 | (0.054) | -0.225 | 3.0 | 0.362 | 0.177 | Green |
| Case 4: | 0.424 | 2.6 | 0.12 | -0.15 | 2.6 | 0.329 | 0.177 | Blue |

Notes:

1. Roughness parameter is fixed at 18º for all cases.

2. Values in parentheses are assumed and kept fixed in the fit.

3. Last column lists the color of lines for these models as plotted in Fig. 8.

**Table. 4.** Hapke model parameters derived from FC narrowband images.

| Filter | $\lambda$ (nm) | SSA | $g$ | $\theta$ | $B_0$ | $h$ | RMS (%) | $A_{geo}$ | $A_{Bond}$ |
|--------|------|-------|--------|------|------|-------|-----|-------|-------|
| F2 | 554 | 0.508 | -0.242 | 17.5 | 1.83 | 0.048 | 4.7 | 0.417 | 0.195 |
| F3 | 749 | 0.544 | -0.245 | 17.3 | 1.72 | 0.044 | 4.5 | 0.439 | 0.211 |
| F4 | 916 | 0.365 | -0.225 | 18.8 | 1.59 | 0.097 | 3.2 | 0.256 | 0.136 |
| F5 | 961 | 0.387 | -0.231 | 17.8 | 1.53 | 0.087 | 3.6 | 0.269 | 0.143 |
| F6 | 828 | 0.470 | -0.248 | 17.7 | 1.46 | 0.065 | 3.9 | 0.343 | 0.177 |
| F7 | 652 | 0.556 | -0.243 | 17.8 | 1.83 | 0.047 | 5.1 | 0.462 | 0.219 |
| F8 | 438 | 0.393 | -0.208 | 18.4 | 1.87 | 0.100 | 4.9 | 0.287 | 0.151 |

**Table. 5.** Hapke model parameters derived from FC narrowband images with $B_0$=1.7 and $h$=0.07 fixed.

| Filter | λ (nm) | SSA | $g$ | $\theta$ | RMS (%) | $A_{\text{geo}}$ | $A_{\text{Bond}}$ |
|--------|--------|-------|--------|------|---------|-------|--------|
| F2 | 554 | 0.500 | -0.229 | 17.7 | 4.7 | 0.376 | 0.195 |
| F3 | 749 | 0.534 | -0.222 | 17.6 | 4.5 | 0.397 | 0.212 |
| F4 | 916 | 0.377 | -0.231 | 18.7 | 3.2 | 0.279 | 0.136 |
| F5 | 961 | 0.391 | -0.230 | 17.8 | 3.6 | 0.289 | 0.143 |
| F6 | 828 | 0.465 | -0.232 | 17.8 | 4.0 | 0.351 | 0.178 |
| F7 | 652 | 0.547 | -0.229 | 18.0 | 5.1 | 0.416 | 0.220 |
| F8 | 438 | 0.408 | -0.233 | 18.1 | 4.9 | 0.305 | 0.151 |

**Table. 6.** Comparisons of the Hapke photometric model parameters of asteroids, comets, Martian satellites, and the Moon that have been modeled with disk-resolved data.

| Object | Type | SSA | $g$ | $\theta$ | $B_0$ | $h$ | $p_v$ | $A_B$ | $\lambda$ (nm) | References |
|---|---|---|---|---|---|---|---|---|---|---|
| Vesta | V | 0.51 | -0.24 | 18 | 1.7 | 0.07 | 0.42 | 0.20 | 554 | This work |
| 103P/Hartley 2 | JFC | 0.036 | -0.46 | 15 | (1.0) | (0.01) | 0.045 | 0.012 | 625 | Li et al. (2013a) |
| 81P/Wild 2 | JFC | 0.038 | -0.52 | 27 | (1.0) | (0.01) | 0.063 | 0.012 | 647 | Li et al. (2009) |
| 9P/Tempel 1 | JFC | 0.039 | -0.49 | 16 | (1.0) | (0.01) | 0.056 | 0.013 | 550 | Li et al. (2007a, 2013b) |
| 19P/Borrelly | JFC | 0.057 | -0.43 | 22 | (1.0) | (0.01) | 0.072 | 0.019 | 660 | Li et al. (2007b) |
| (253) Mathilde | C | 0.035 | -0.25 | 19 | 3.18 | 0.074 | 0.041 | 0.013 | 700 | Clark et al. (1999) |
| Deimos | C | 0.079 | -0.29 | 16.4 | 1.65 | 0.068 | 0.067 | 0.027 | 540 | Thomas et al. (1996) |
| Phobos | C | 0.07 | -0.08 | 22 | 4 | 0.05 | 0.056 | 0.021 | 540 | Simonelli et al. (1998) |
| Phoebe | C | 0.068 | -0.24 | 31 | 3.4 | 0.038 | 0.81 | 0.2 | 480 | Simonelli et al. (1999) |
| Average C | - | 0.037 | -0.47 | 20 | 1.03 | 0.025 | 0.049 | 0.012 | - | Helfenstein and Veverka (1989) |
| (1) Ceres | G | 0.07 | -0.4 | 44 | 1.58 | 0.06 | 0.088 | 0.02 | 555 | Li et al. (2006); Helfensten and Veverka (1989) |
| (433) Eros | S | 0.33 | -0.25 | 28 | 1.4 | 0.010 | 0.23 | 0.092 | 550 | Li et al. (2004) |
| (243) Ida | S | 0.22 | -0.33 | 18 | 1.53 | 0.020 | 0.21 | 0.07 | 560 | Helfenstein et al. (1996) |
| Dactyl | S | 0.21 | -0.33 | 23 | (1.53) | (0.020) | 0.2 | 0.065 | 560 | Helfenstein et al. (1996) |
| (951) Gaspra | S | 0.36 | -0.18 | 29 | 1.63 | 0.06 | 0.22 | 0.11 | 560 | Helfenstein et al. (1994) |
| (25143) Itokawa | S | 0.42 | -0.35 | 26 | 0.87 | 0.01 | 0.33 | 0.14 | 1570 | Kitazato et al. (2008) |
| (5535) Annefrank | S | - | - | - | - | - | 0.24 | - | 647 | |
| Average S | - | 0.23 | -0.27 | 20 | 1.6 | 0.08 | 0.18 | 0.08 | - | Helfenstein and Veverka (1989) |
| (2867) Steins | E | 0.57 | -0.30 | 28 | 0.60 | 0.062 | 0.39 | 0.24 | 630 | Spjuth et al. (2012) |
| (9969) Braille | Q | - | - | - | - | - | 0.34 | - | 540 | Buratti et al. (2004) |
| Moon, Highland | - | 0.51, 0.48 | -0.34, -0.32 | 20 | 1 | 0.06, 0.15 | 0.28 | 0.13 | 750 | Hillier et al. (1999) Kennelly et al., (2010) |
| Moon, Maria | - | 0.33 | -0.23, -0.21 | 20 | 1 | 0.07, 0.15 | 0.17 | 0.07 | 750 | |

Note: Values in parentheses are assumed.

**Figure captions**

Fig 1 - Disk-integrated phase function Vesta measured from Dawn FC images acquired during approach. The diamond symbols at phase angles >23º are from Dawn FC images through CLEAR filter. The photometric measurements have been corrected to a common cross-section of Vesta with an equivalent radius of 260 km, and corrected from the effective wavelength of CLEAR filter (699 nm) to *V*-band wavelength (550 nm) using the spectral slope of Vesta's spectrum measured from the ground. The small red dots at phase angles below ~25º are from ground-based data (Lagerkvist and Magnusson, 1995). The blue crosses at phase angles ~53º are from Rosetta OSIRIS measurement (Fornasier et al., 2011). The solid line is the best-fit IAU H-G phase function model based on measurements at phase angles lower than 80º. The dashed line is a best-fit Hapke model with parameters listed in the figure, as discussed in Section 6.2. At phase angles <80º, the disk-integrated phase function of Vesta can be well described by the IAU H-G phase function model. The Hapke model describes the phase function well at all phase angles of our observations.

Fig 2 - Density plots of the *I/F* data points extracted from images acquired from ON19 to Survey through F2 filter (554 nm). Panel (a) shows *I/F* with respect to phase angle; panel (b) with respect to incidence angle; and panel (c) the emission angle. The colors in the plot correspond to the number density of data points in a linear stretch from zero the maximum density as represented by the color bar at the bottom of the figure. The group of data points with very low *I/F* at incidence angles between 80º and 100º, emission angles between 25 and 65º, and phase angles between 50º and 65º should be due to scattered light in the shadowed areas and night side near or beyond the terminator of Vesta at high-latitude, most likely at northern mid-
to high-latitude. We discarded those data in our modeling with $I/F<0.012$, $i>90º$, or $e>90º$.

Fig 3 - Density plots of the disk-resolved $I/F$ data to show the slight trend between the scattering
geometry of disk-resolved photometric data and latitude. Panel (a) shows that there is a slight
trend between phase angle and latitude. Panel (b) shows that there is a strong trend between
incidence angle and latitude. Panel (c) shows that there is almost no trend between emission
angle and latitude, but the majority of data have low to moderate emission angles. The colors
in the plot correspond to the number density of data points in a linear stretch from zero the
maximum density as represented by the color bar at the bottom of the figure.

Fig 4 - The Minnaert model results for the $I/F$ data from F2 filter (554 nm, Fig. 2). Panel (a)
shows the modeled Minnaert albedo with diamond symbols. The statistic error bars are
smaller than the size of symbols. The solid line is a linear fit in magnitude with a zero-phase
angle Minnaert albedo (almost equivalent to normal albedo for Vesta excluding the opposition
effect) $0.27\pm0.03$. Panel (b) shows the modeled Minnaert $k$ parameter with diamond symbols
and error bars. The solid line is a linear fit to the $k$ parameter. The RMS of 6.7% refers to the
total model RMS for all parameters shown in this figure.

Fig 5 - The model scatter characteristics of the Minnaert model shown in Fig. 4. Panel a shows
the modeled $I/F$ plotted against measured $I/F$. The RMS is 6.7%, and the linear correlation
between measurement and model is 0.985. Panel b shows the ratio between measured $I/F$ and
modeled $I/F$ with respect to scattering geometry. The scatter along incidence angle, emission
angle, and phase angle are about $\pm0.2$. There does not appear to be any systematic trend with
respect to scattering geometry. However, at ~40º phase angle and higher, the model scatter
significantly increases.

Fig 6 - Minnaert parameters with respect to wavelengths. The Minnaert *A* shown in panel a is the best-fit values at zero-phase angle, almost equivalent to normal albedo for Vesta (excluding opposition effect). The broad trend of normal albedo, except for the point at 699 nm derived from CLEAR filter (cross symbol in all panels), is consistent with the spectral shape of Vesta in the visible wavelengths. Panel b shows the slope of Minnaert *A* parameter with respect to phase angle, equivalent to a measure of the phase slope, as a function of wavelength. Within the 1-μm band (last two points) and in blue wavelength (438 nm) the phase slopes appear to be slightly steeper. Panel c shows the Minnaert *k* parameter extrapolated to zero-phase angle. There does not appear to be any obvious trend with respect to wavelength, and the parameters are all within a narrow range of 0.532 and 0.545. Panel d is the slope of Minnaert *k* parameter with respect to phase angle plotted as a function of wavelength. Again, no trend with wavelength is evident.

Fig 7 - Comparisons between the best-fit Hapke models from disk-resolved data and the observed disk-integrated phase function of Vesta through clear filter. Panel b is the zoom-in view of panel a within 20° phase angles. The various models are plotted in color lines, and their parameters are listed in Table. 2. The photometric measurements from various sources shown in Fig. 1 are plotted as dots in this figure. All these models can fit the disk-integrated phase function almost equally well.

Fig 8 - Hapke model fit to the disk-integrated phase function of Vesta. Panel b is a zoom-in view within 20° phase angle. The model parameters for various lines are listed in Table. 3. The photometric measurements from various sources shown in Fig. 1 are plotted as dots in this figure and used for model. All these models can fit the disk-integrated phase function almost equally well. The largest discrepancy between various models occurs within phase

angles <20º, but still less than 0.2 mag, comparable to the amplitude of Vesta's rotational lightcurve.

Fig 9 - The best-fit Hapke model parameters and derived quantities for all FC color filters, plotted with respect to wavelength. In all panels, filled blue triangles are model parameters with $B_0$=1.7 and $h$=0.07 fixed, and listed in Table. 5; and open red squares are model results with all five parameters set free, as listed in Table. 4. Panel i shows the correlations between $h$ and SSA.

Fig 10 - The model scatter characteristics for the best-fit Hapke model for $I/F$ data of F2 filter (554 nm, Fig. 2). Panel a is the measured $I/F$ vs. modeled $I/F$. Panel b shows the ratio between measured $I/F$ and modeled $I/F$ with respect to scattering geometry. There is no systematic trend for the model scatter with geometry. The model scatter is about ±0.2 with a few points at emission angles between 75º and 80º showing scatter up to +0.7.

Fig 11 - The comparisons between the modeled geometric albedos that we derived from FC data and SMASS-II data and HST data. HST data points were photometrically calibrated (Li et al., 2011). The error bars for HST data points represent the range of the rotational variations of Vesta's geometric albedo. The SMASS-II spectrum was scaled to match the HST data points. For both cases of modeling that we performed, the geometric albedos of Vesta that we derived are systematically higher than HST measurements in 500 − 750 nm wavelength range by up to 25% if $B_0$ and $h$ are set free, and up to 15% if $B_0$ is set to 1.7 and $h$ 0.07.

Fig 12 - The error estimate for $B_0$ as an example to demonstrate our approach to derive the model uncertainties for all parameters. Panel a shows the $\chi^2$ with respect to $B_0$ in the range of 0-3.0, with other parameters adjusted accordingly to reach local minima. Panel b shows the corresponding geometric albedo for various values of $B_0$ in the solid line. The diamond symbol is the best-fit $B_0$ and geometric albedo. The dashed line marks the ±20% geometric albedo boundaries and the corresponding values of $B_0$. Panel c shows the disk-integrated phase function corresponding to the various values of $B_0$ (thin black lines) and the best-fit phase function (thick red line).

Fig 13 -    Photometric scans and the models along the equivalent photometric equator and mirror meridian for $I/F$ data extracted from RC3 and RC3b at two phase angles, ~37º and ~8º, with a pixel sizes of ~0.51 km/pix at Vesta. The symbols are the average $I/F$ data along the photometric scans, averaged over the whole surface. The dotted vertical lines mark the geometric range where $i$<80º and $e$<80º that we performed our fit. The solid lines are the profiles predicted by the best-fit Hapke model. The long-dashed lines are parameterless Akimov model. The short-dashed lines are the Minnaert model. The dash-dot lines are the Lommel-Seeliger model. The dash-triple-dot lines are the LS-Lambertian model. Lommel-Seeliger model is consistent with the photometric equator scan but not the mirror meridian scan. The LS-Lambertian model fits the data at phase angles <60º along photometric equator, but not at higher phase angles or along the mirror meridian. The Minnaert model produces bad fit near limb. The Hapke model and Akimov model appear to be the best description of all cases.

Fig 14 -    Photometrically corrected radiance factor mosaics obtained through F3 filter (749 nm) from RC3b (upper panel) and RC3 (lower panel), shown in a sinusoidal projection. The grid lines are spaced by 30º for both longitude and latitude, with a center longitude of 180º. The photometric correction was performed with the best-fit photometric parameters with $B_0$=1.7 and $h$=0.07 fixed (Table. 5) to the standard geometry with $i$=30º, $e$=0º, and $\alpha$=30º.

The average phase angle of RC3b images is ~10º, and that of RC3 images is ~37º. The areas with *i*>50º or *e*>75º were trimmed, leaving blanks in the northern latitude and near the south pole in both mosaics. Two maps are displayed with the same linear brightness stretch. The map from higher phase angle (RC3) appears to show more residual topography and more photometric correction artifacts than the map from lower phase angle (RC3b). Seams are slightly visible in both maps, but more obvious in the RC3 map at higher phase angle. These mosaics suggest that the overall quality of photometric correction with the Hapke model parameters is acceptable, although more work is needed to improve it.

Fig 15 -   The average radiance factor of Vesta at 554 nm along longitude (panel a) and latitude (panel b), measured from the photometrically corrected mosaic derived from RC3b data (Fig. 14). The dotted line is the global average. Note that panel a may not represent the true average radiance factor of Vesta along longitude because of the incomplete coverage of the mosaic in the northern hemisphere. Also it is fundamentally different from a rotational lightcurve, which represents the projected-area-weighted average reflectance of a whole hemisphere centered at particular longitudes. Overall, the dark areas on Vesta distribute concentrate in longitude from 60º to 190º, and north of -30º latitude. The reflectance of the southern hemisphere is dominated by Rheasilvia basin, which is brighter than the global average by ~10%, and than the northern hemisphere by ~18%.

Fig 16 -   The histograms of photometrically corrected *I/F* as derived from the mosaics shown in Fig. 14 show single-peaked distribution of reflectance on Vesta. The distribution from RC3b (solid line), which has a relatively low phase angle, shows a small shoulder on the low-reflectance side, while the one from RC3 (dashed line) does not. The slightly wider base of the distribution from RC3 than that from RC3b is due possibly to more artifacts introduced

by photometric correction in the former than the latter, as discussed in the text. With the incomplete surface coverage of these maps and imperfect photometric correction, it is uncertain whether the shoulder is real or not. But it is relatively certain that the albedo distribution of Vesta is different from that of the Moon, which shows obvious dichotomy between mare and highlands.

Fig 17 -    Bolometric Bond albedo map of Vesta as generated by combining the photometrically corrected mosaics from all FC color filter.

Fig 18 -    Panel a shows radiance factor ($I/F$) under scattering geometry of $i$=30º, $e$=0º, and phase=30º, as a function of SSA, as predicted by Hapke model with the best-fit parameters for Vesta through F2 filter (554 nm), as listed in Table. 5. The horizontal dotted line marks the maximum $I/F$ observed for Vesta at ~0.5 km/pix. The triangle marks the modeled disk-averaged SSA for Vesta at this wavelength, 0.52. The dashed line between SSA 0.35 and 0.65 is a linear fit to the curve in this segment. Although Vesta's reflectance is non-linear with respect to SSA, within ±30% of the average SSA, the reflectance is approximately linear to the SSA within 4%. At much higher reflectance (2x), the non-linearity must be considered. Panel b is a density plot showing the fraction of multiple scattering at 554 nm, as predicted by Hapke model with the best-fit model parameters. The colors in the plot correspond to the number density of data points in a linear stretch from zero the maximum density as represented by the color bar at the bottom of the figure. For most data about 20% to 30% of scattered light is multiply scattered.

Fig 19 -    Panel a shows the best-fit model phase functions of Vesta (as listed in Table. 5). Panel b shows the ratio of the phase functions, normalized to unity at zero phase angle, with respect to the one from F3 filter at 749 nm. And panel c is the zoom in of panel b at phase

angles <30º.  Ratios of less than unity indicate that the phase function is steeper than that at 749 nm.  Overall the ratio is within 5% from unity at phase angles lower than 25º (accessible from the ground), and up to 15% within 120º phase angle.  The steepest phase functions are near the center of the 1-µm absorption at 917 nm, and at wavelength 438 nm.

**Fig. 1.**

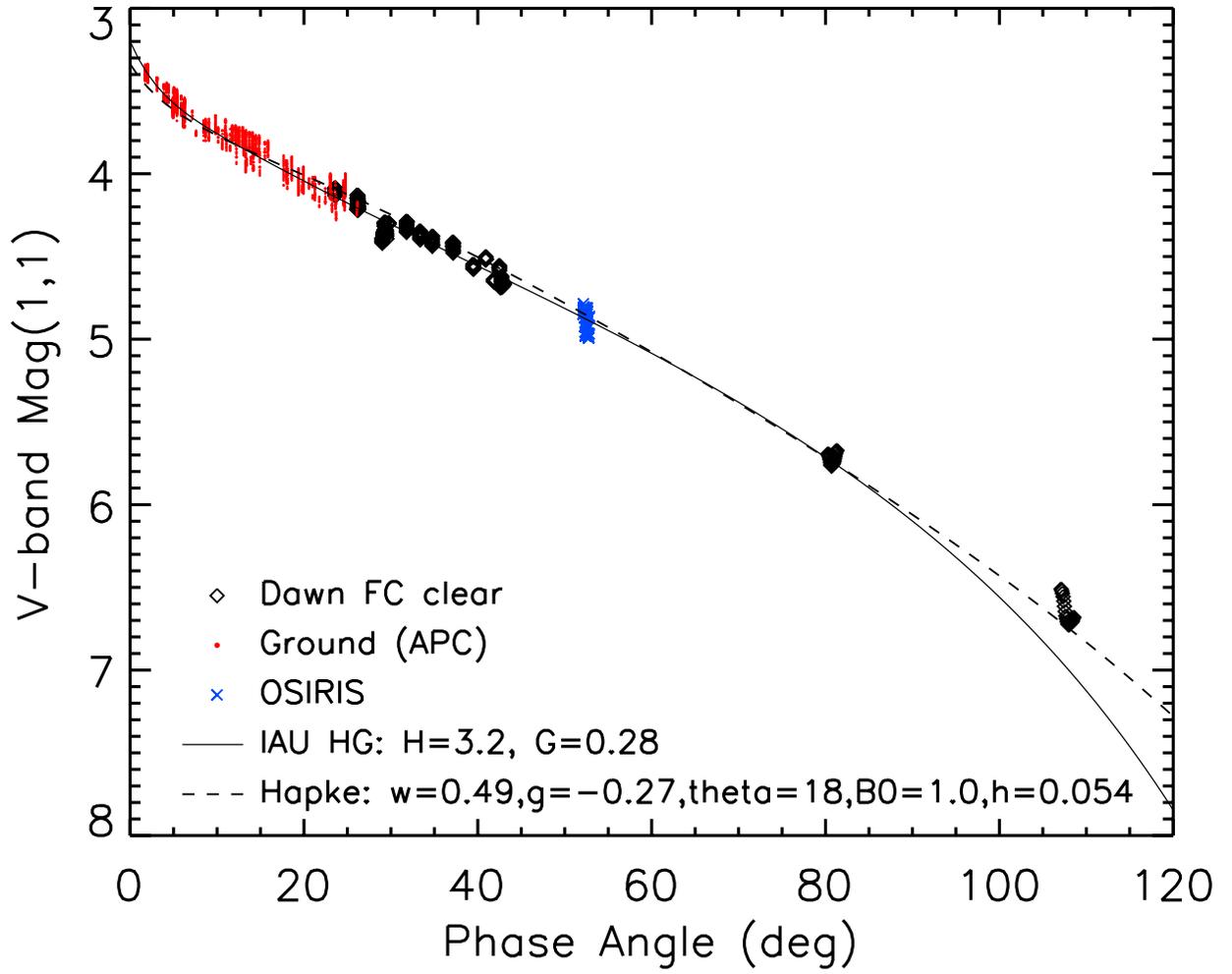

**Fig. 2.**

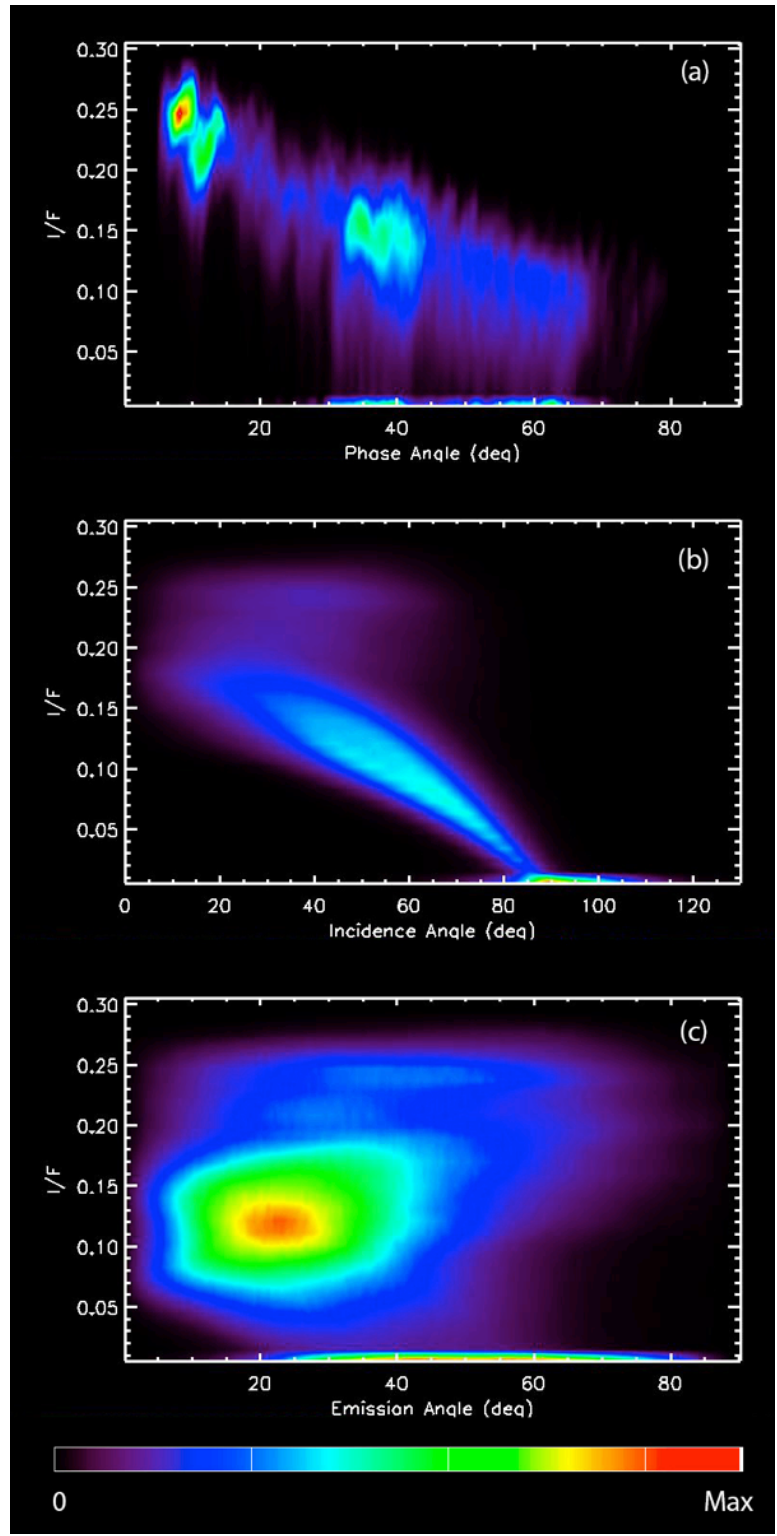

**Fig. 3.**

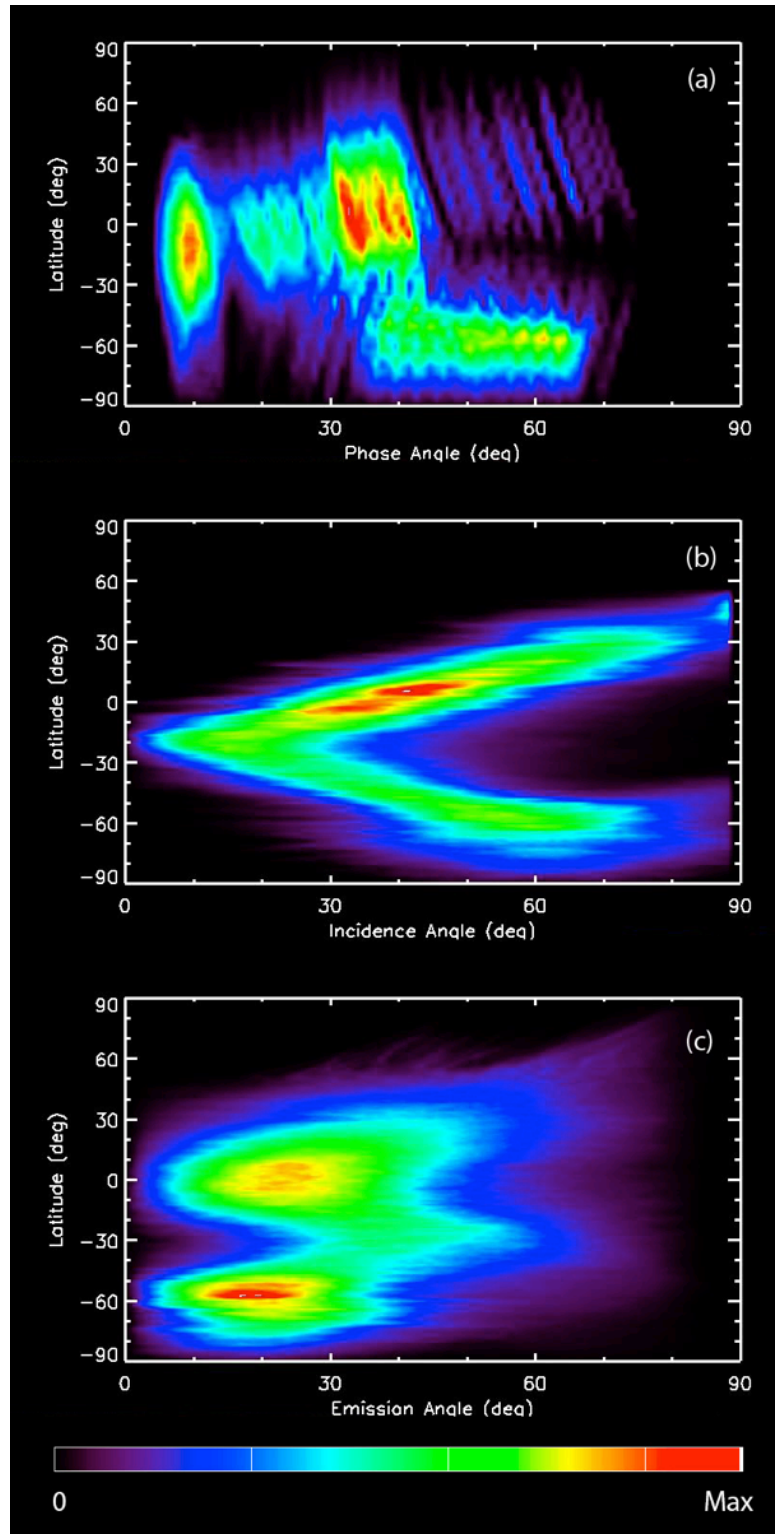

**Fig. 4.**

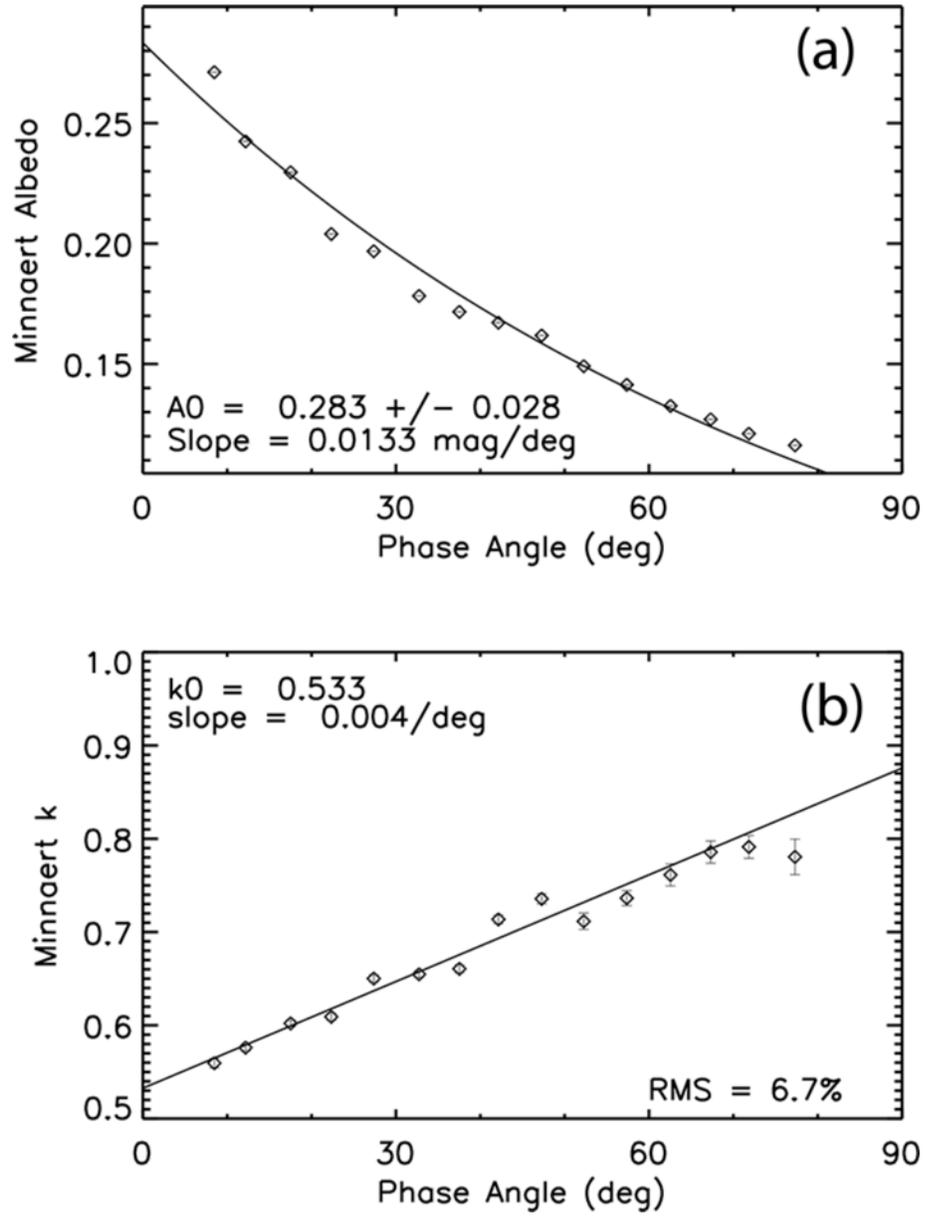

**Fig. 5.**

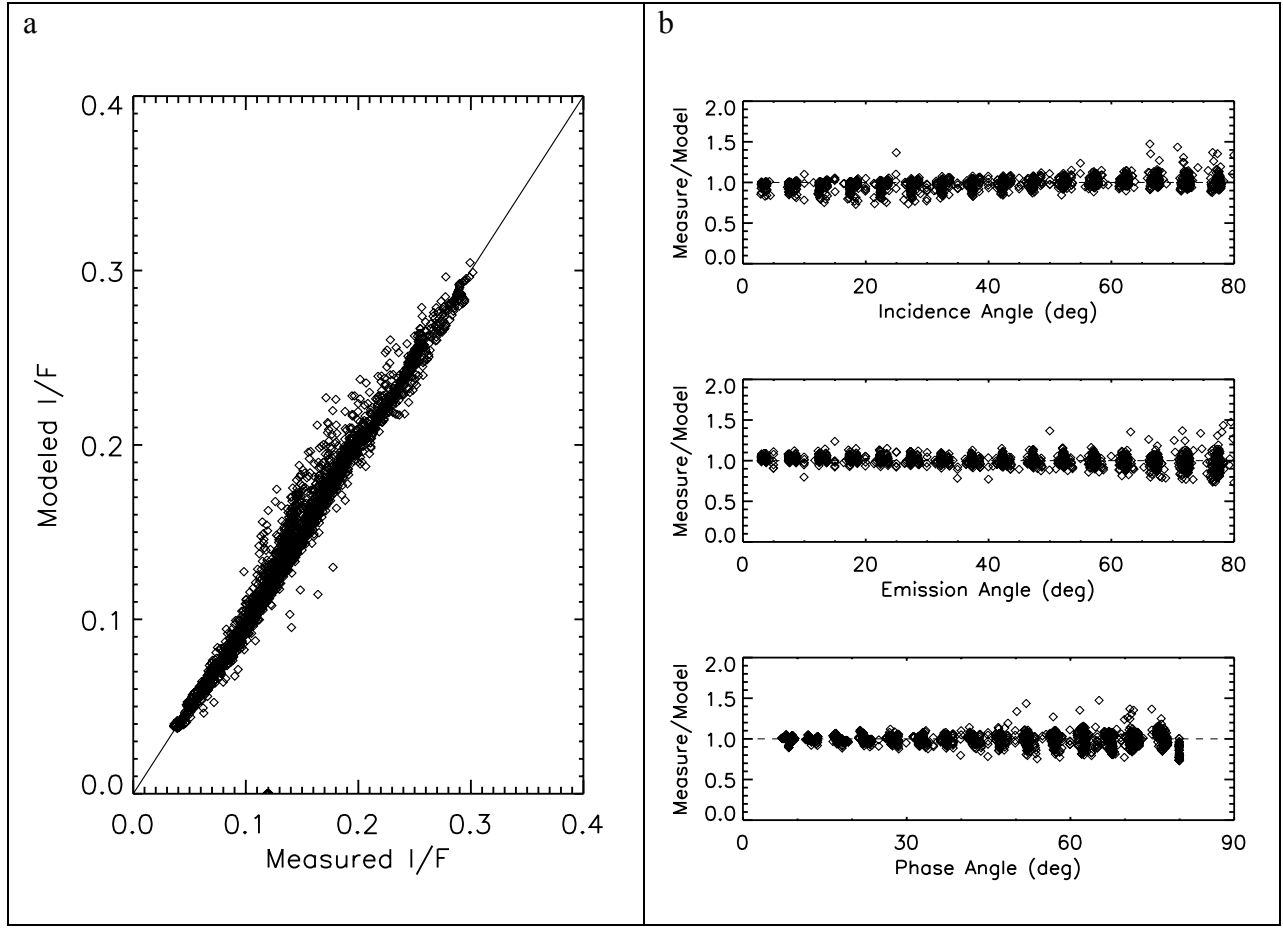

**Fig. 6.**

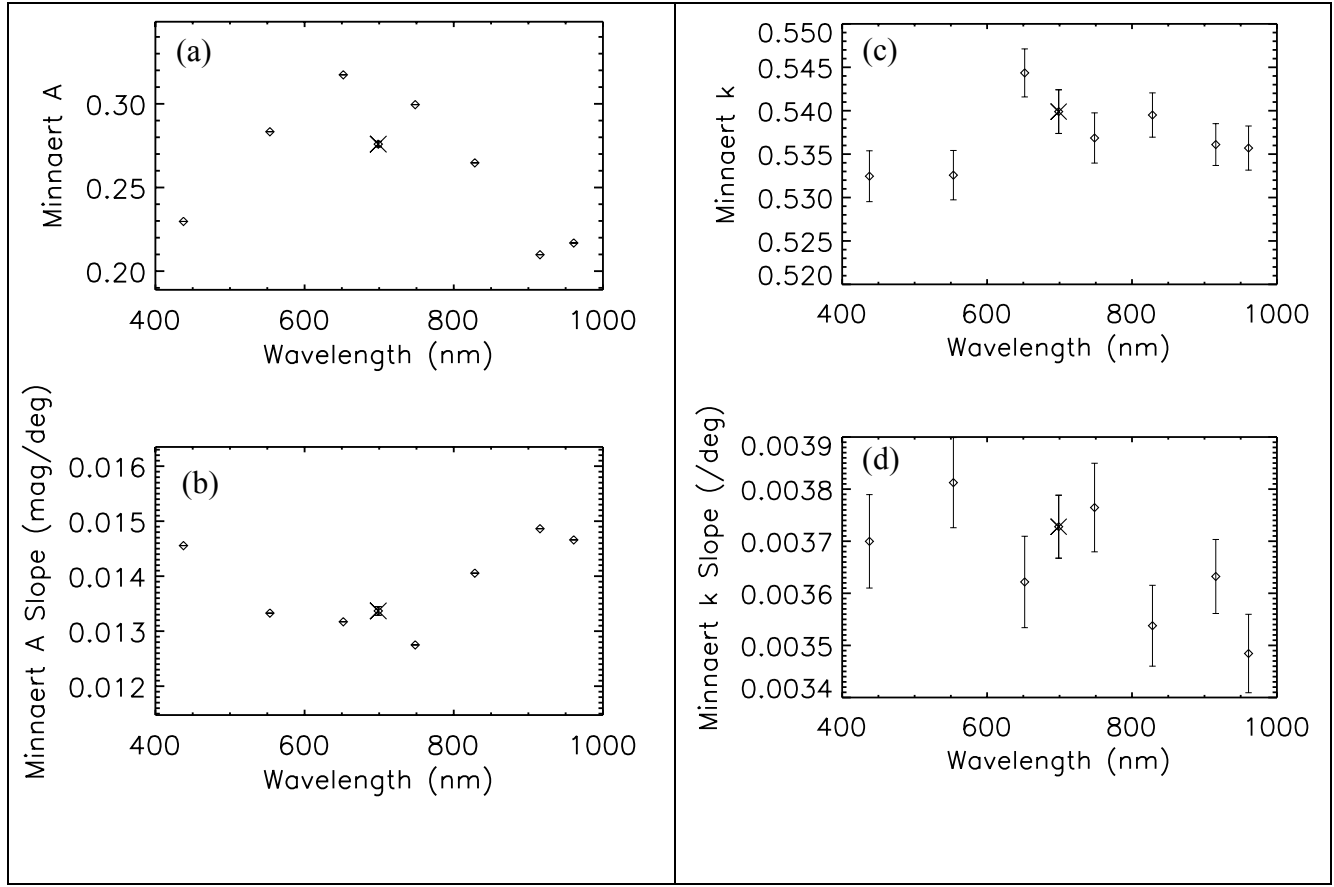

**Fig. 7.**

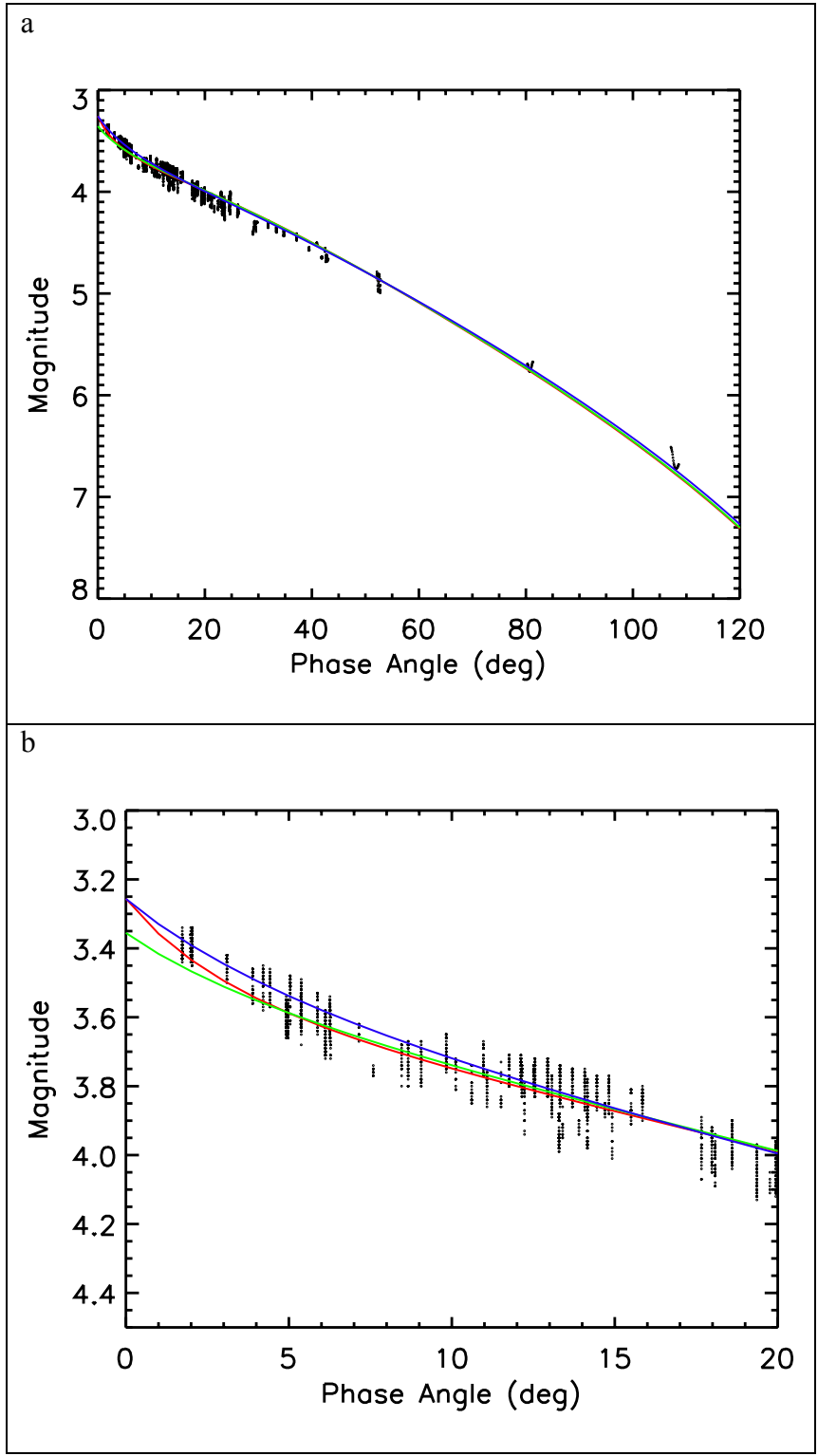

**Fig. 8.**

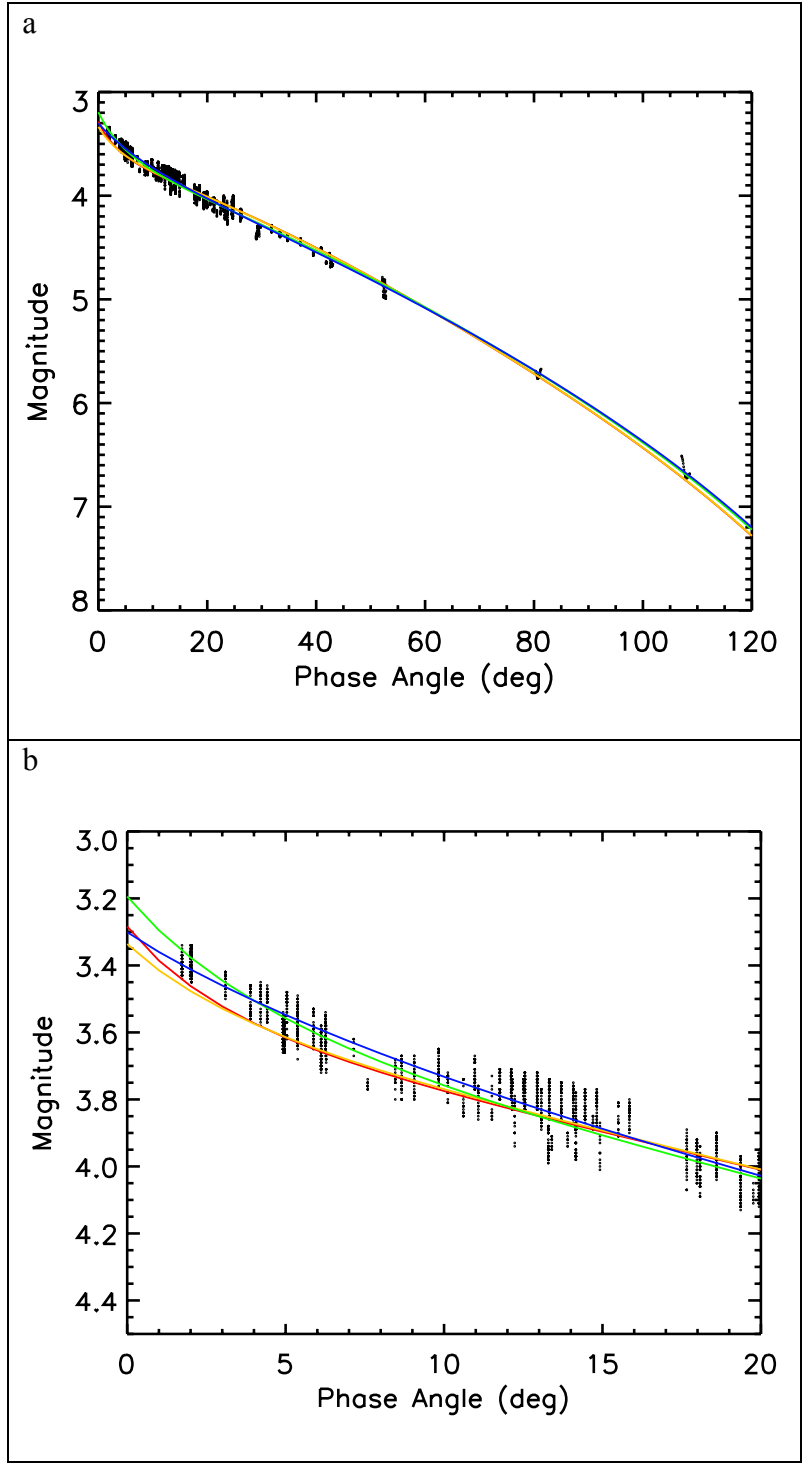

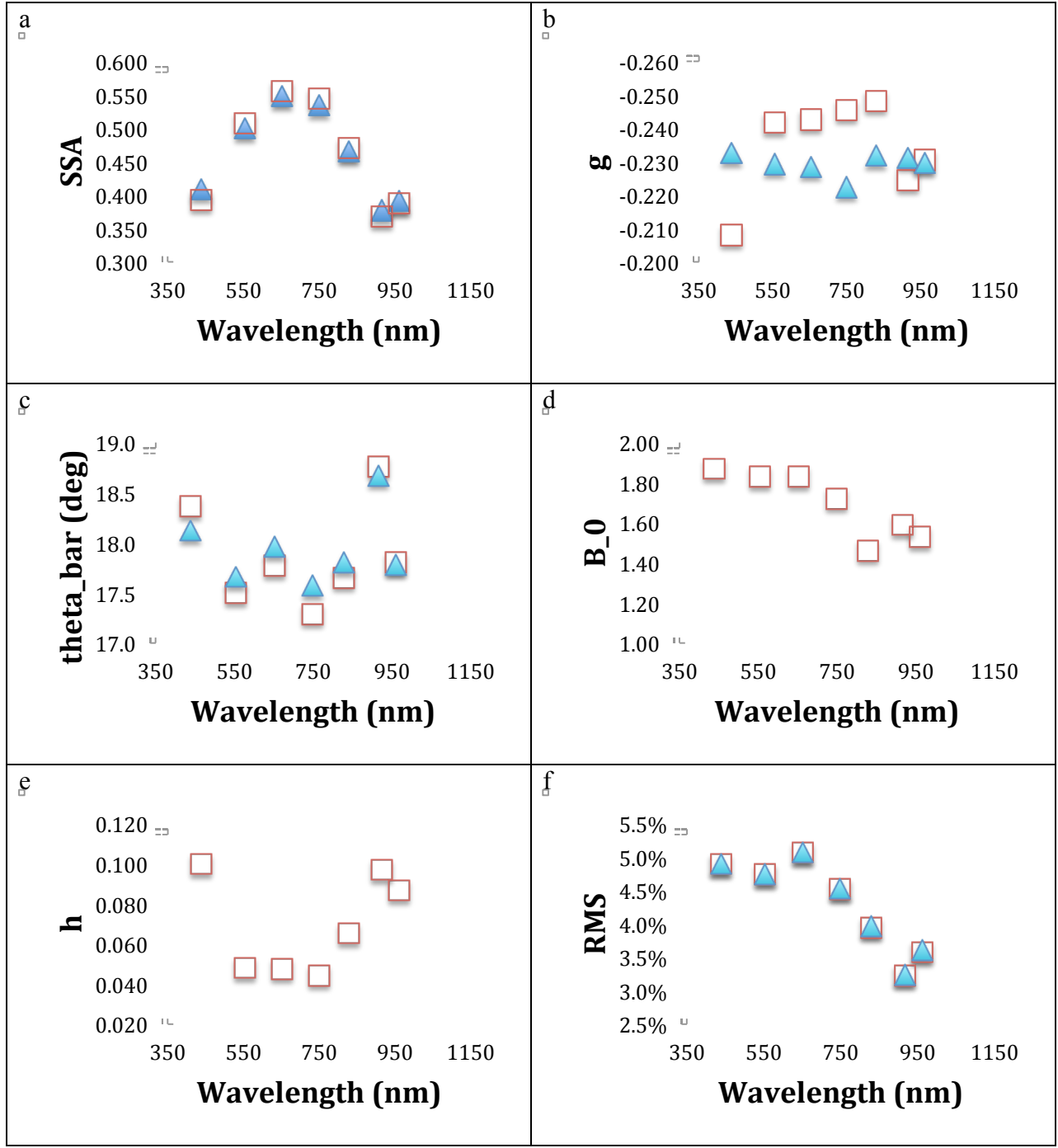

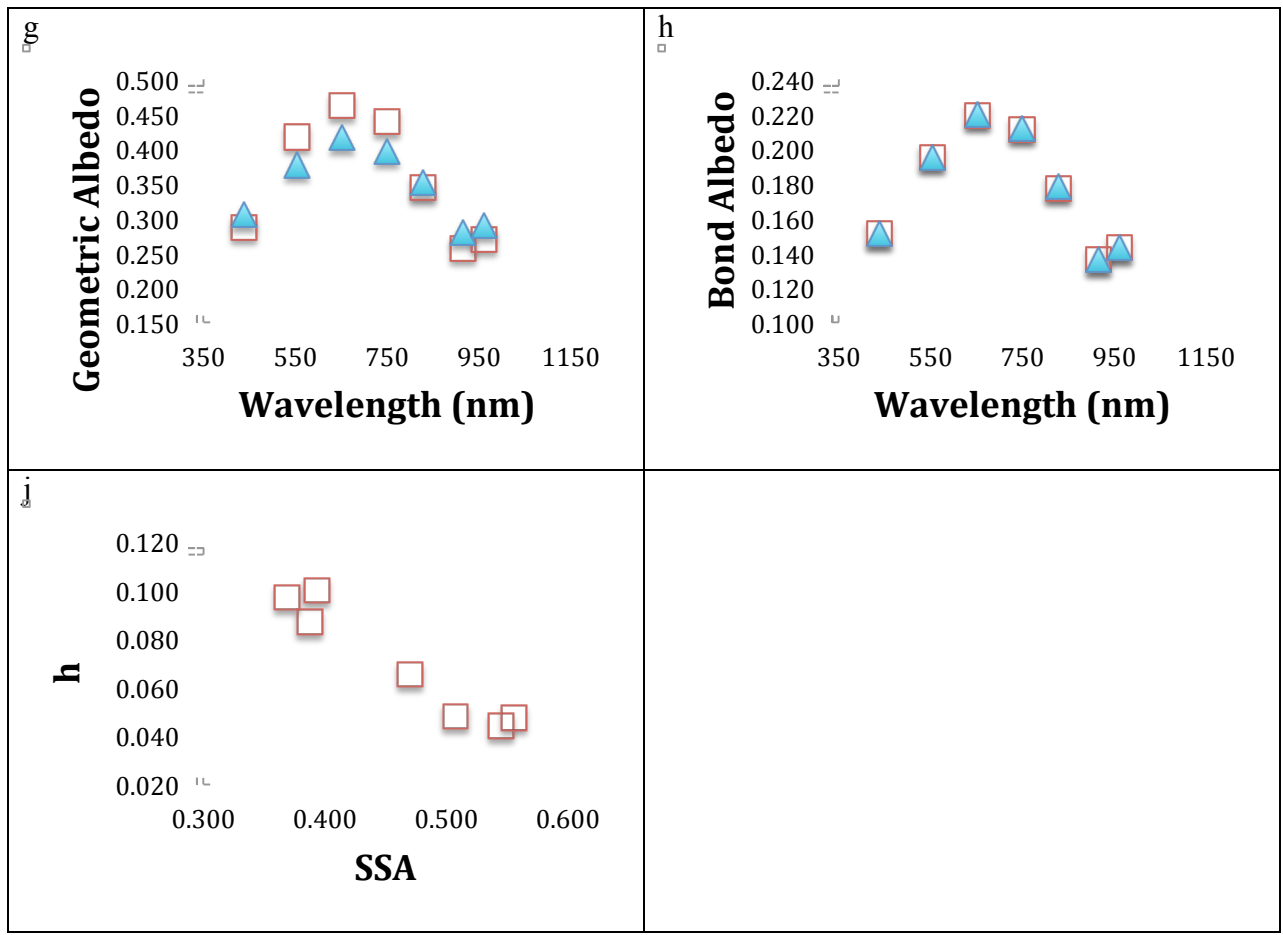

**Fig. 10.**

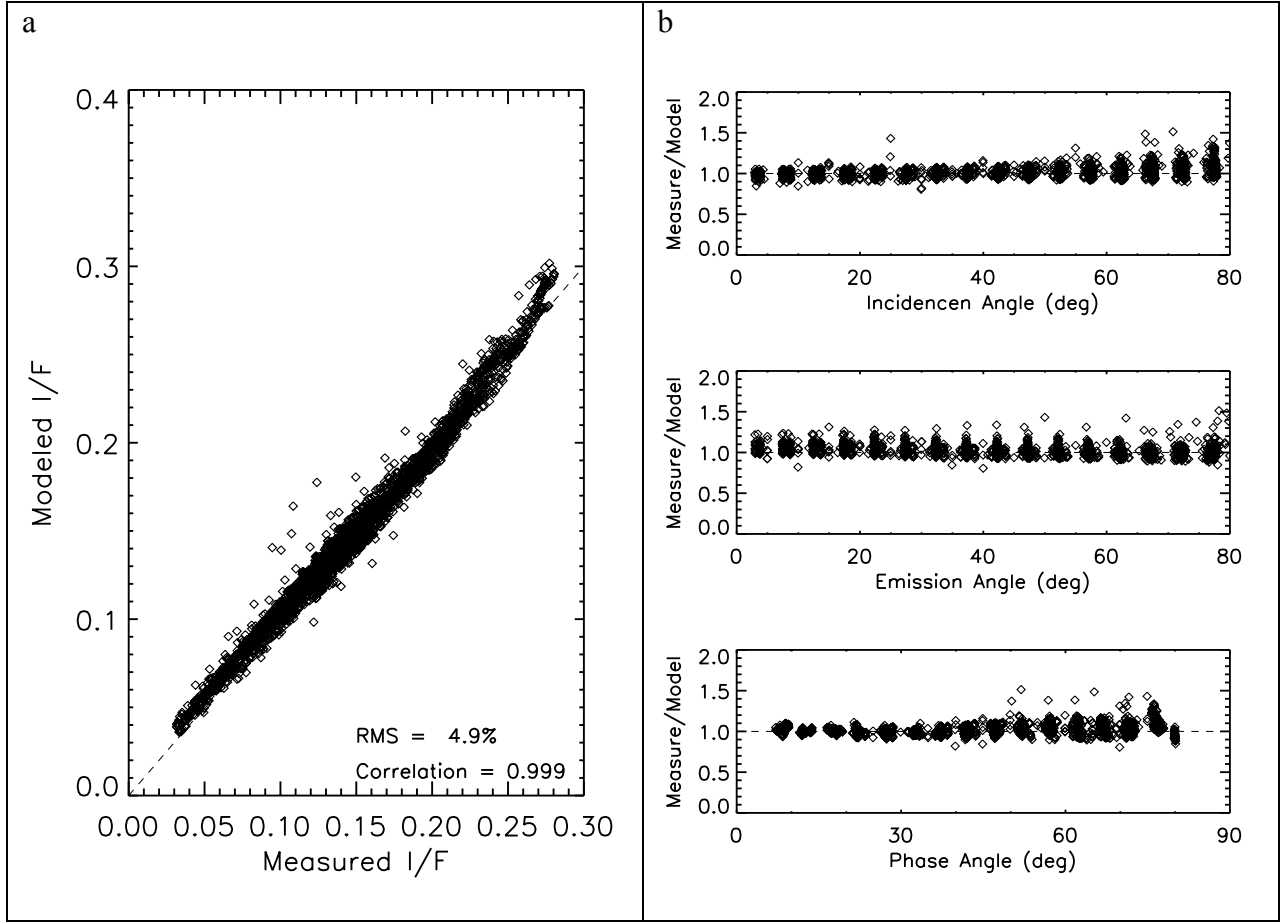

**Fig. 11.**

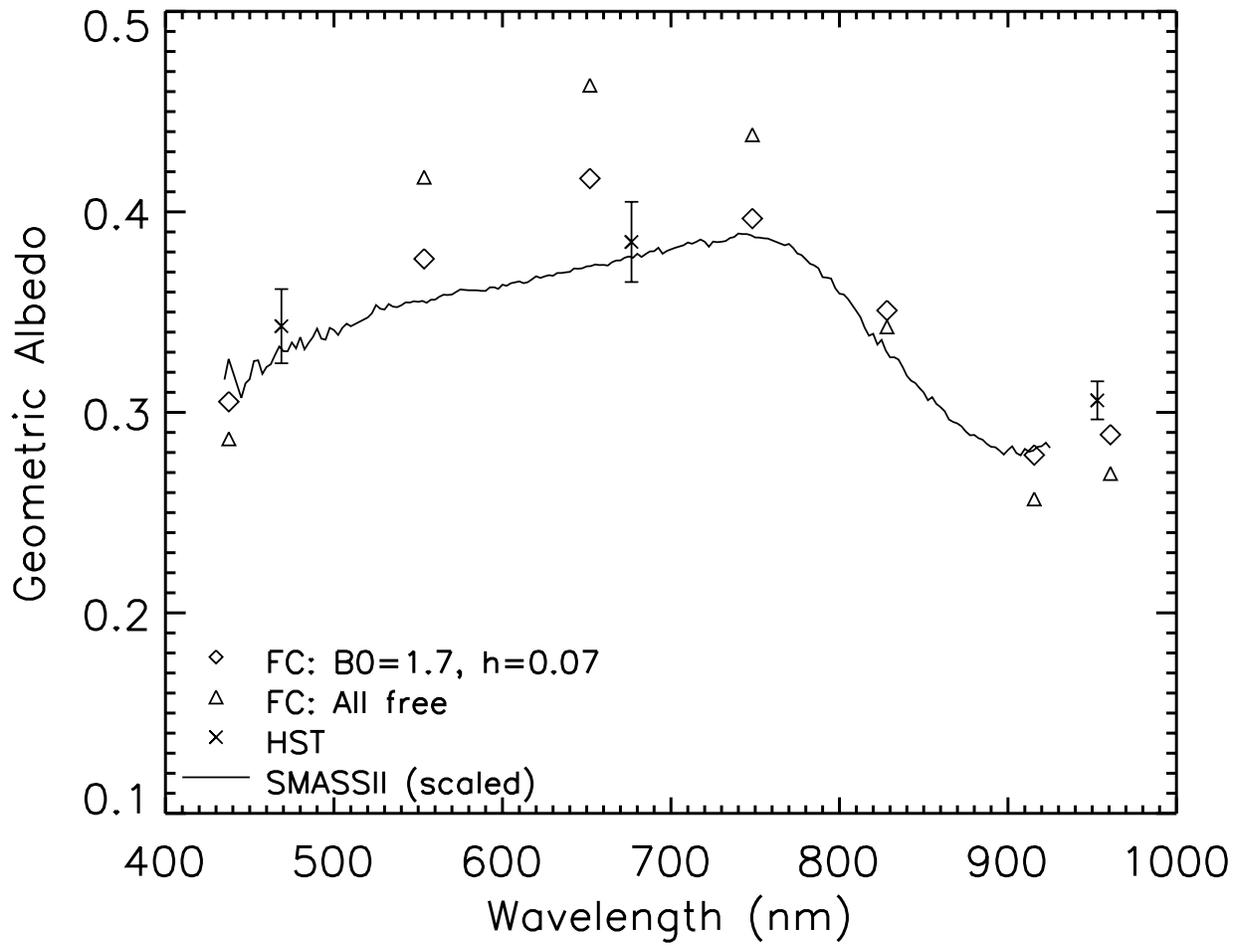

**Fig. 12.**

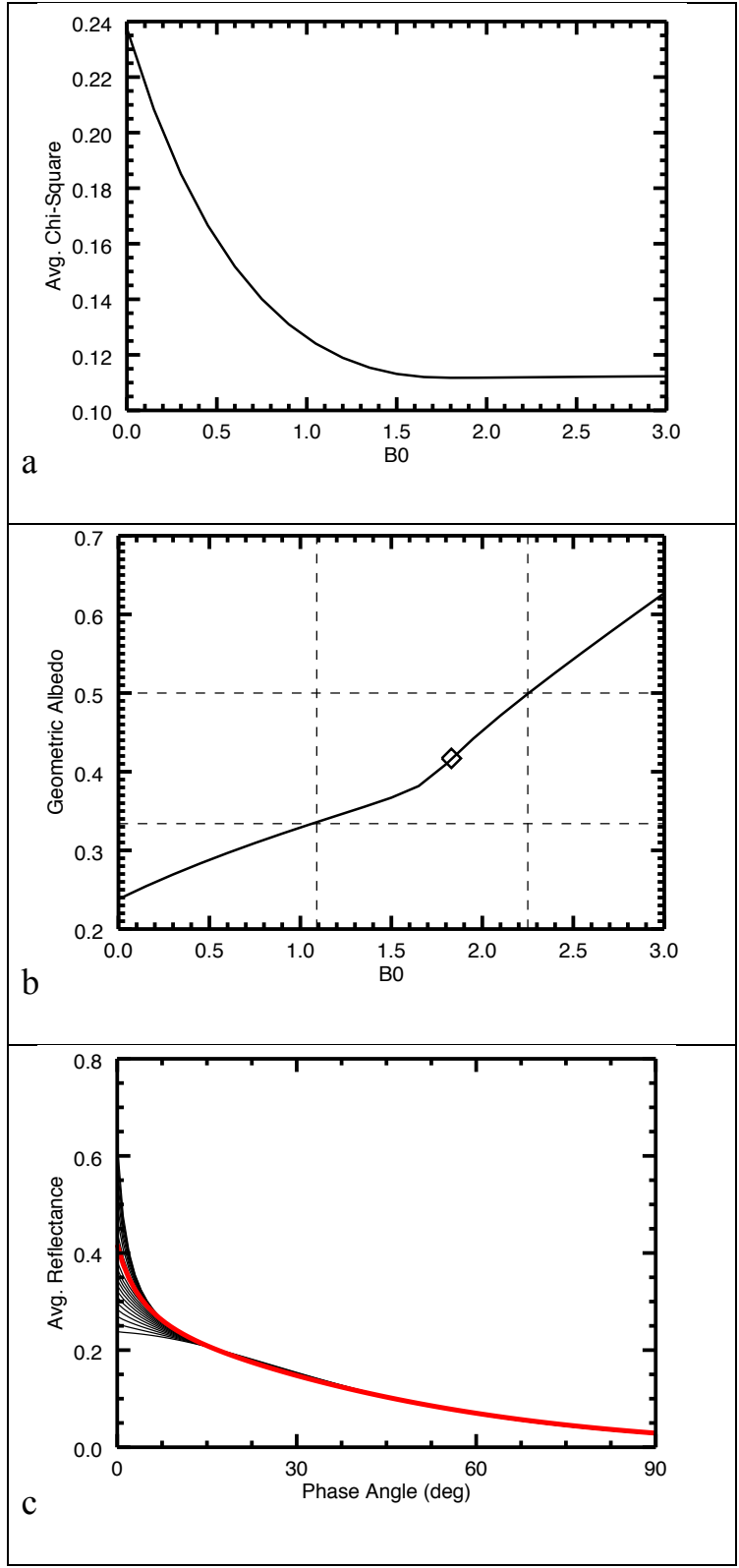

**Fig. 13.**

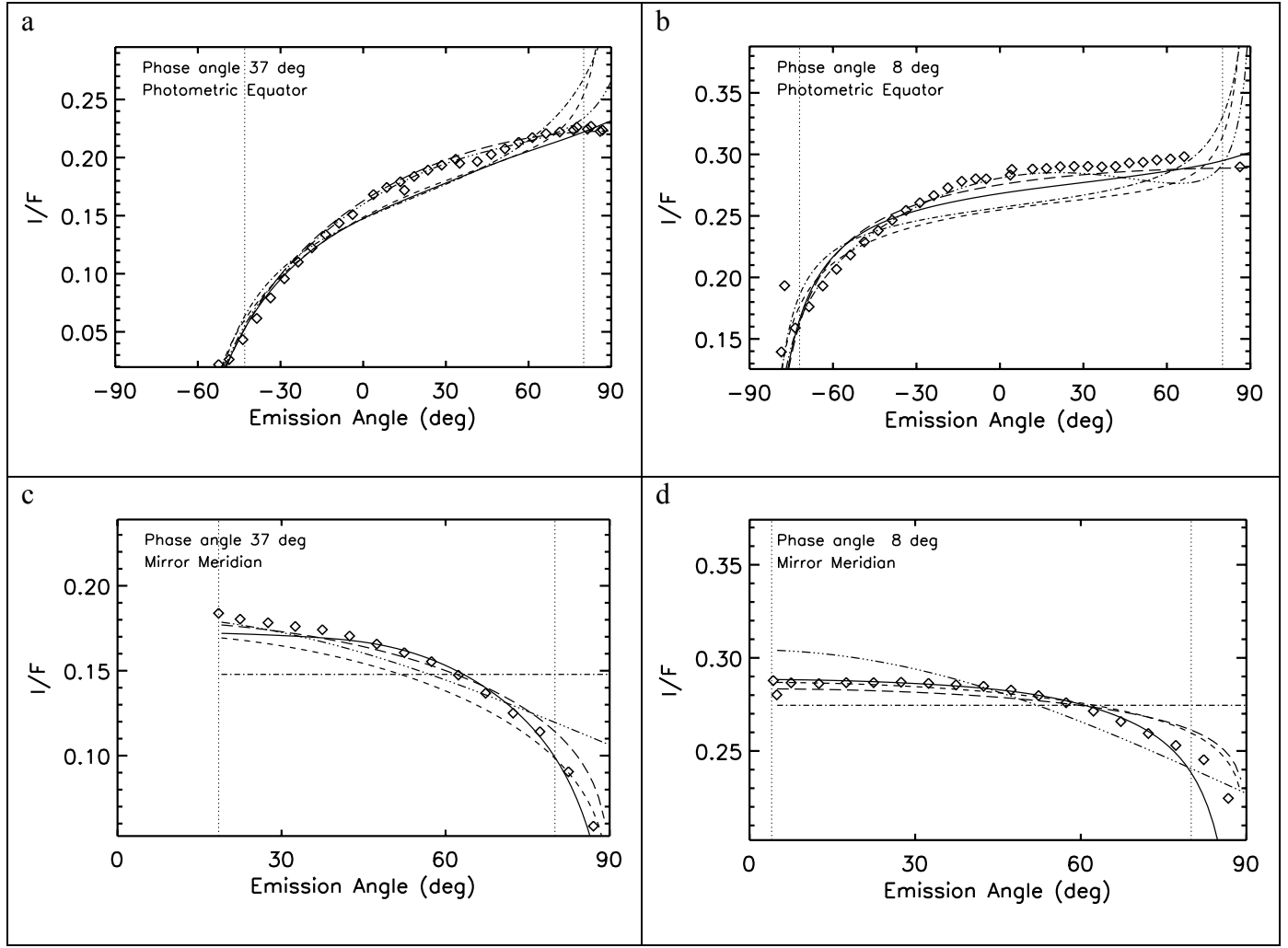

**Fig. 14.**

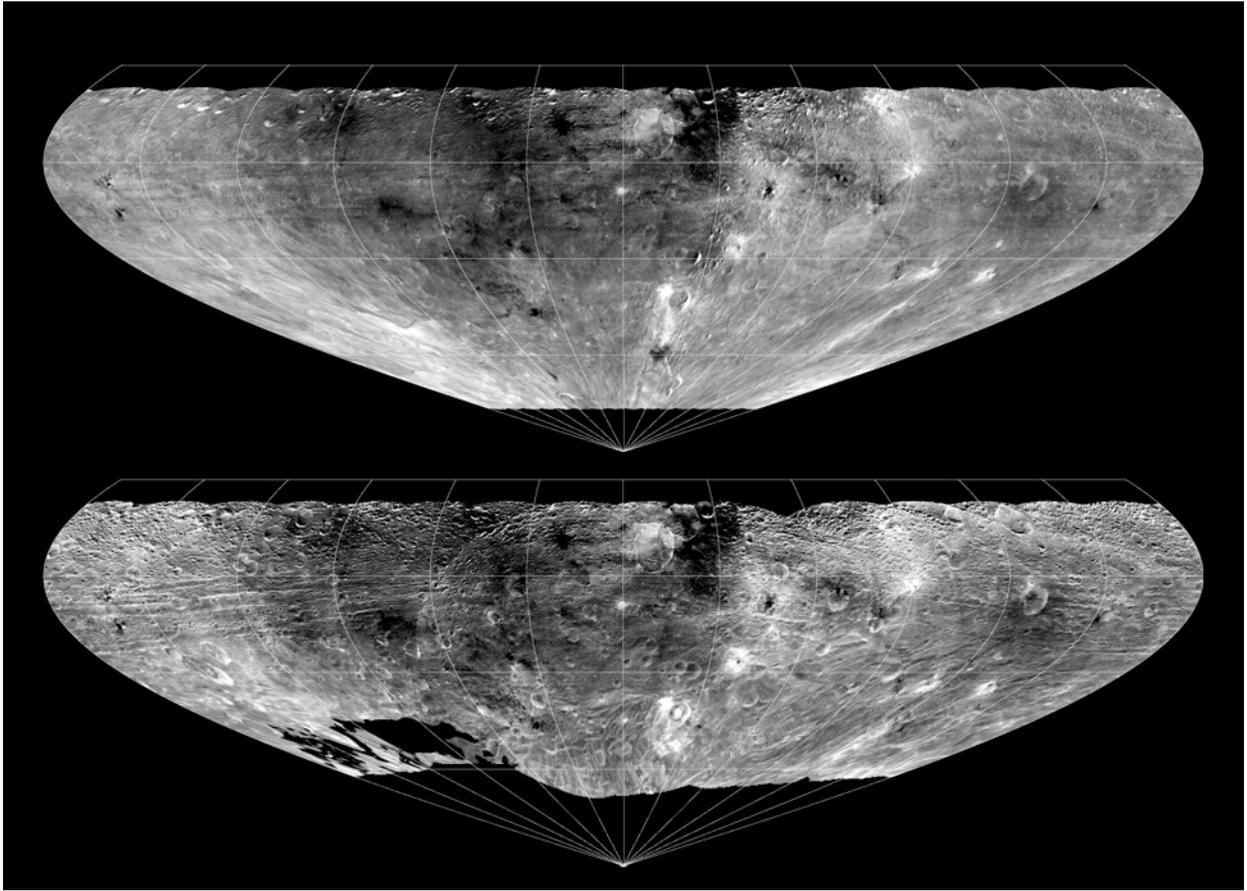

**Fig. 15.**

(a)

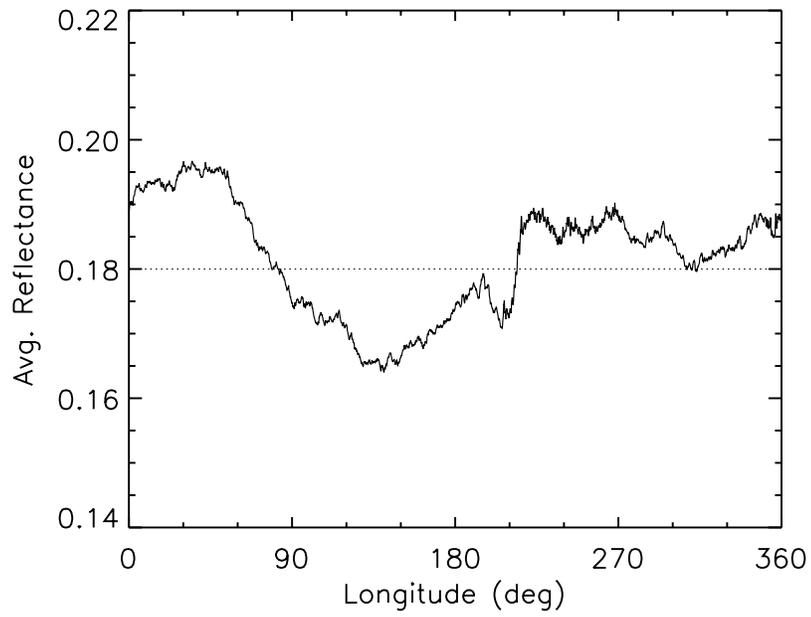

(b)

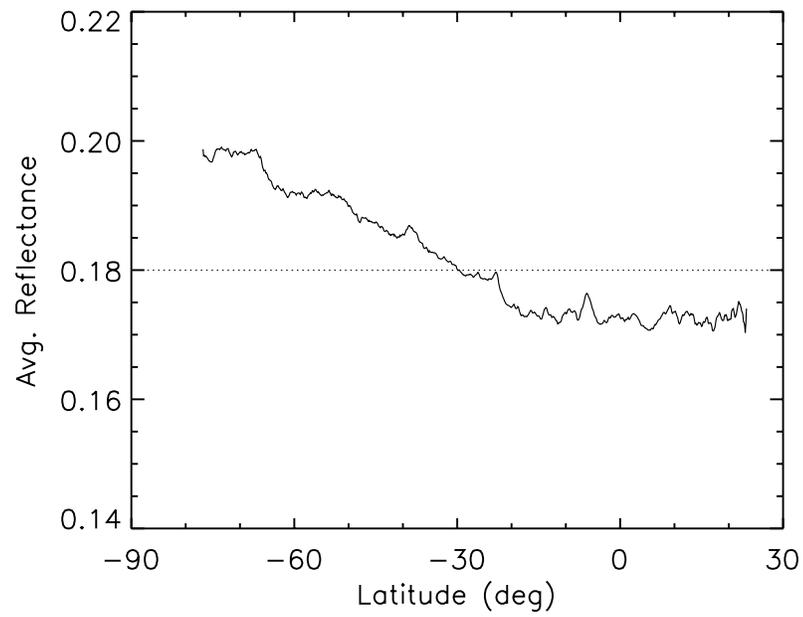

**Fig. 16.**

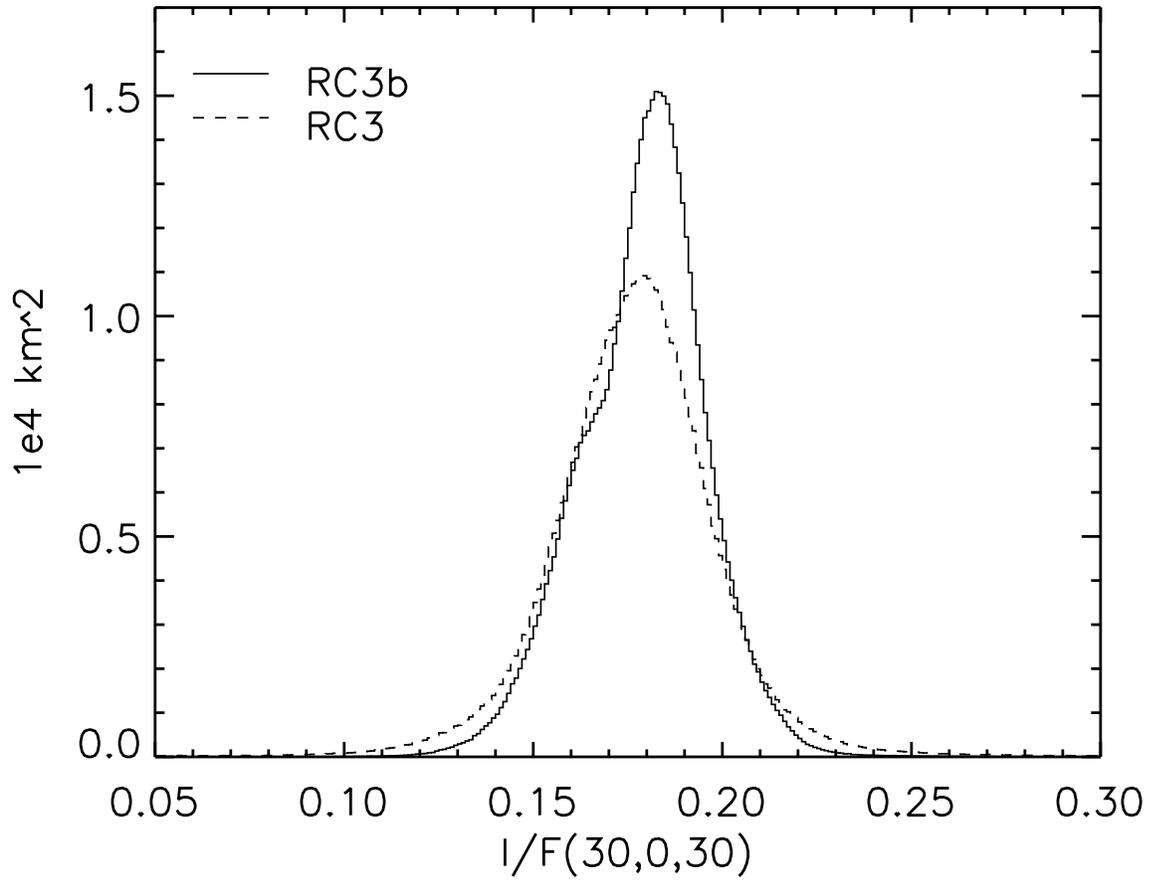

**Fig. 17.**

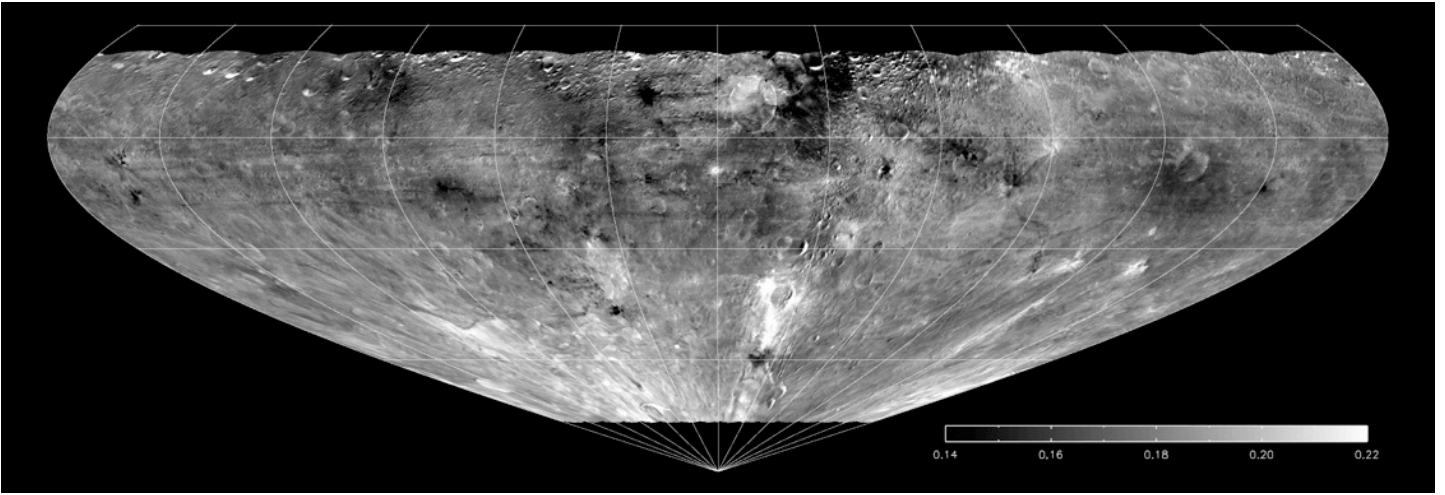

**Fig. 18.**

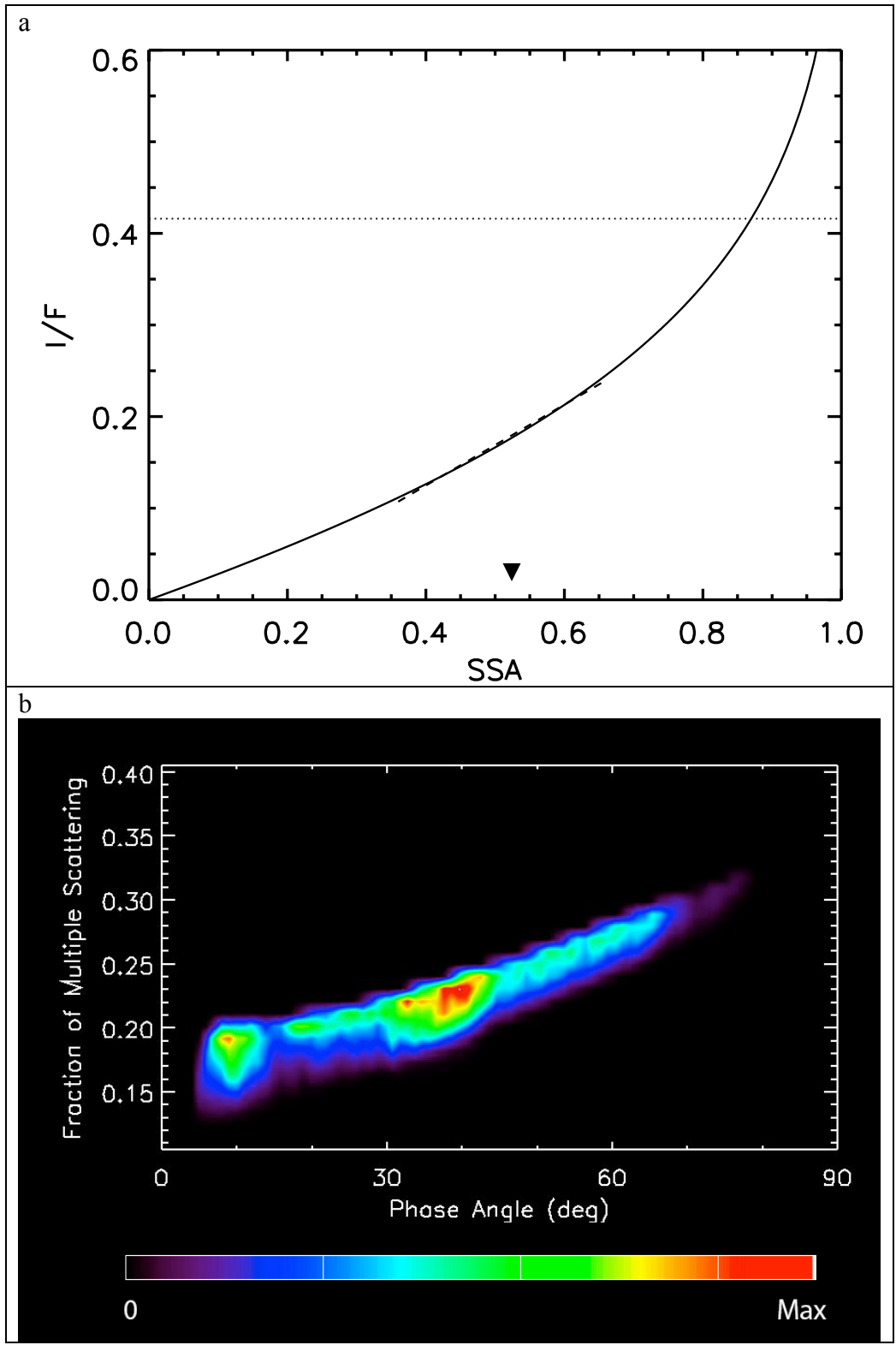

**Fig. 19.**

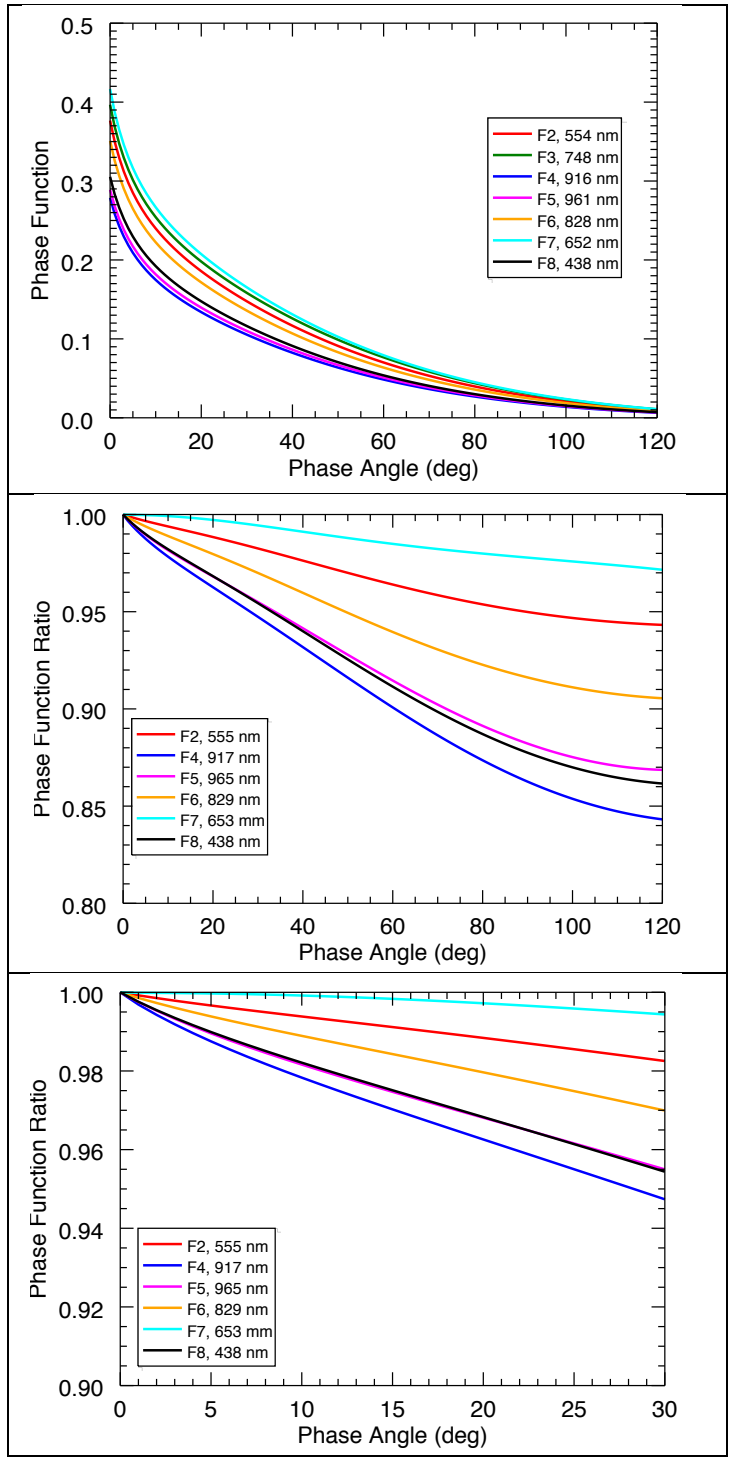